\documentstyle[11pt,epsf]{article}

\def\conlim{\mathop{\longrightarrow}\limits_{\scriptscriptstyle R \to \infty
, T = R/2}}

\newcommand{\eq}{\begin{equation}}
\newcommand{\en}{\end{equation}}
\newcommand{\eqa}{\begin{eqnarray}}
\newcommand{\ena}{\end{eqnarray}}

\begin{document}

\hbox{}
\noindent {31 May 1996} \hfill HU Berlin--IEP--96/16\\
\mbox{} \hfill FSU-SCRI-96-47\\
\mbox{} \hfill SWAT/96/107\\
\begin{center}
\vspace*{1.5cm}
\renewcommand{\thefootnote}{\fnsymbol{footnote}}
\setcounter{footnote}{0}
{\LARGE Compact U(1) Lattice Gauge--Higgs Theory with Monopole Suppression}
\footnote{Work partly supported by the HCM Network
``Computational Particle Physics" grant CHRX-CT92-0051} \\
\vspace*{1.5cm}
{\large
Balasubramanian Krishnan $\mbox{}^1$\footnote{Lise Meitner Postdoctoral
Research Fellow sponsored by FWF under Projects M078-PHY and M212-PHY;
Current address: Datametrics Systems Corporation, 12150 E. Monument Drive,
Suite 300, Fairfax, VA 22033, USA},
U.M.~Heller        $\mbox{}^2$,
V.K.~Mitrjushkin $\mbox{}^{3,4}$
\footnote{Supported
by the Deutsche Forschungsgemeinschaft under research grant Mu 932/1-3;
Permanent address: Joint Institute for Nuclear Research, Dubna, Russia},
M.~M\"uller-Preussker $\mbox{}^4$,
}\\
\vspace*{0.7cm}
{\normalsize
$\mbox{}^1$ {\em Technische Universit\"at Wien, Institut f\"ur Kernphysik,
A-1040 Wien, Austria}\\
$\mbox{}^2$ {\em SCRI, Florida State University, Tallahassee,
FL 32306-4052, USA} \\
$\mbox{}^3$ {\em University of Wales, Swansea, U.K.} \\
$\mbox{}^4$ {\em Humboldt-Universit\"{a}t, Institut f\"ur Physik,
Berlin, Germany}}\\
\vspace*{2cm}
{\bf Abstract}
\end{center}

We investigate a model of a U(1)--Higgs theory on the lattice with compact
gauge fields but completely suppressed (elementary) monopoles. We study the
model at two values of the quartic Higgs self--coupling, a strong coupling,
$\lambda = 3.0$, and a weak coupling, $\lambda=0.01$. We map out the phase
diagrams and find that the monopole suppression eliminated the confined
phase of the standard lattice model at strong gauge coupling. We perform a
detailed analysis of the static potential and study the mass spectrum
in the Coulomb and Higgs phases for three values of the gauge coupling.
We also probe the existence of a scalar bosonium to the extent that our
data allow and conclude that further investigations are required
in the Coulomb phase.
\newpage
\section{Introduction}

For many years the compact formulation of QED on a lattice \cite{wil} has
been advocated as  a preferable alternative to the standard non--compact
one. There are at least two arguments in favor of quantum electrodynamics
with a compact gauge group \cite{pol1,pol2}. The abelian gauge group of QED
has to be compact if it appears from some non--abelian, unified gauge
group. For compact groups the charge is quantized, which is not the case
for the non--compact theory. Both formulations are supposed to have
identical perturbative series in the continuum limit, but may lead to
completely different infrared behavior on a nonperturbative level.

If we couple U(1) gauge fields to dynamical scalar fields -- scalar QED --
we obtain the simplest prototype of a unified theory.  The phase structure
of the corresponding compact lattice theory and the nature of the vortex
string excitations have  been intensively studied over the years (see, for
instance, [\ref{pd1}--\ref{pd2}]). Scalar QED shows confinement at  strong
coupling similar to QCD and  Coulombic behavior at weak coupling
corresponding to the real world of electromagnetic phenomena.  For
sufficiently large values of the parameter $v$ -- the classical position
of the minimum of the Higgs potential -- the theory exhibits the
conventional Higgs behavior. Nevertheless, the nature of the Coulomb--Higgs
transition, in comparison with the chiral transition, as well as the
spectral content of the theory in the Coulomb and the Higgs phases deserve
further studies. In the non--compact case with non--linear Higgs field this
has been done recently \cite{bfkk}. The transition turned out to be most
likely second order and the critical behavior $\lambda \phi^4$ like, {\it
i.e.,} trivial in the continuum limit.

When scalars in the compact theory are in the fundamental representation
there is no phase boundary that completely separates the confined and Higgs
phases. The confined and Higgs regions in the phase diagram remain
analytically connected. This observation, first made for lattice models
with frozen out radial modes  of the Higgs field \cite{br,FS},  was later
confirmed for more realistic theories with radially varied Higgs fields and
for different gauge groups. It has led to the conjecture of a principle of
complementarity: a confining theory of fermions and gauge bosons can be
analyzed as if a dynamical Higgs phenomenon takes place or as if the gauge
symmetry is unbroken \cite{drs}.

Provided all results of lattice calculations can be applied to continuum
physics, the existence of the Higgs--confinement phase  could entail rather
strong physical consequences (see, e.g., \cite{drs,fas}). However, at
finite lattice spacing, lattice theories suffer from lattice artifacts.
These can be simply scaling violations, {\it i.e.,} effects from irrelevant
operators that vanish in the continuum limit, as is believed to be the case
for lattice QCD. Or they can consist of field configurations on the lattice
which have no continuum counterpart but strongly influence the infrared
properties of the lattice theory. The confinement phase in the compact
abelian lattice theory  is a non--physical phenomenon traced back to the
particular discretization of gauge fields on the lattice. The
confinement--Coulomb phase transition is due to the condensation of lattice
monopoles, {\it i.e.,} artifacts \cite{dgt}. It is straightforward to
modify Wilson's theory such that monopoles are removed from the functional
measure in the euclidean functional integral \cite{bss,BMM}.

In this paper we want to discuss the abelian gauge--Higgs theory on the
lattice with compact U(1) gauge fields in a {\it modified} version,
with monopoles suppressed. As shown in \cite{bss,BMM} for the pure $~U(1)~$
gauge theory the phase transition  to the confinement phase disappears at
real bare coupling leaving only a  unique Coulomb phase \cite{BMM}.  The
suppression of monopoles in the U(1) theory with staggered fermions
entails the disappearance of the chiral transition in the zero--mass limit
(at least, in the quenched approximation) \cite{hmm1}.

In the theory with monopoles suppressed the Higgs phase is not anymore part
of the Higgs--confinement phase, similar to the non--compact formulation,
as a confinement phase does not exist. The (conjectured) complementarity
principle therefore can not be applied in this case. It is interesting to
see to what extent this influences the physics (spectrum, etc.) in the
Higgs phase.

Dealing with the modified theory also has some technical advantages in
comparison with the standard compact version when considering the Coulomb
phase. Since monopoles are completely suppressed, problems related to their
existence disappear. For instance, it is easier to identify the lowest
photonic states in plaquette--plaquette correlators, because of the smaller
overlap with higher energy states with photon quantum numbers \cite{BMM}.

Here we discuss the complete phase structure of this model including also
imaginary, {\it i.e.,} unphysical bare coupling values. Within the Higgs
and the Coulomb phases we try to identify and to interpret the lowest lying
gauge and Higgs field excitations. This is a technically important and, as
we shall see, difficult task.

In the second section we formulate our model and introduce all notations.
The third section is devoted to the study of the phase structure of the
modified theory and its comparison with that of the standard one. In the
fourth section we investigate the behavior of the heavy charge potential,
the screening energy and the Fredenhagen--Marcu order parameter in the
vicinity of the Higgs phase transition. The fifth section is devoted to the
study of the spectrum in the Higgs and Coulomb phases. The last section is
reserved for conclusions.

\section{The Model}

The compact lattice U(1) Higgs model with a scalar field $\phi$  of charge
one is defined by the action
\begin{eqnarray}\label{com_act}
S_{UH}& = &\beta\sum_{\Box} \left[ 1- \frac{1}{2} (U_{\Box} + U_{\Box}^{*})
 \right]
 -\kappa \sum_{x}\sum_{\mu=1}^{4}(\phi_{x}^{*}U_{x,\mu}\phi_{x+\hat{\mu}} + c.c.)
\nonumber \\
  &  & + \lambda \sum_{x} (
\phi_{x}^{*}\phi_{x} - 1)^{2} + \sum_{x}\phi_{x}^{*}\phi_{x}
\end{eqnarray}
where the first term corresponds to the Wilson action $S_{W}$ with
$U_{\Box}$ as the product of the link variables $U_{l} \in$ U(1) around a
plaquette and the  remaining terms correspond to the $\lambda \varphi^{4}$
continuum action  with minimal coupling to the gauge fields,
\begin{equation}
S_{\varphi^4} = \int d^{4}x [ |D_{\mu}\varphi_{x}|^{2}
+\tilde{m}^{2}|\varphi_{x}|^{2} + \tilde{\lambda} |\varphi_{x}|^{4} ].
\end{equation}
The relationship between the parameters of the continuum and lattice
actions are:
\begin{equation}
a\varphi_{x} = \phi_{x}\sqrt{\kappa}, \quad \tilde{\lambda}=
\lambda/\kappa^{2}, \quad (a\tilde{m})^{2}=(1-2\lambda - 8\kappa)/\kappa\,.
\end{equation}
$\phi$ can be decomposed into a radial and angular part as follows:
\begin{equation}\label{angles}
\phi_{x} = \rho_{x}\sigma_{x};\quad \quad \rho_{x} \in {\cal R},\;\;
\sigma_{x} \in {\rm U(1)}.
\end{equation}
The Wilson action $S_{W}$ which is the first term in eq. (\ref{com_act})
may be given in terms of the plaquette angles $\theta_{x;\mu\nu}\in
[-\pi,\pi)$
\begin{equation}\label{wil_act}
S_{W} = \beta \sum_{x;\mu\nu}(1- \cos\theta_{x;\mu\nu}).
\end{equation}
where, given the link angles $\theta_{x,\mu} \in [-\pi,\pi)$ with
$U_{x,\mu} = {\rm e}^{i\theta_{x,\mu}}$, the plaquette
angle is expressed as $\theta_{x;\mu\nu} = \theta_{x,\mu}
+ \theta_{x +\hat{\mu},\nu} - \theta_{x +\hat{\nu},\mu} - \theta_{x,\nu}$.
The physical flux $\bar{\theta}_{x;\mu\nu} \in [-\pi,\pi)$ is then defined
as
\begin{equation}
\theta_{x;\mu\nu} = \bar{\theta}_{x;\mu\nu} + 2 \pi n_{x;\mu\nu}
\end{equation}
where $n_{x;\mu\nu}$ measures the number of Dirac strings.
The modified
Wilson action supplemented by a monopole term $S_{MA}$ \cite{bss,BMM}
is of the form
\begin{equation}
S_{MA} = S_{W} + \gamma \sum_{x;\rho} |M_{x;\rho}|
\end{equation}
where $\gamma$ is a chemical potential controlling the density of monopoles
and $M_{x;\rho}$ measures the net monopole flux out of the 3D cube
orthogonal to the direction $\rho$:
\begin{equation}
2\pi M_{x;\rho} = \frac{1}{2}\epsilon_{\rho\sigma\mu\nu}
(\bar{\theta}_{x+\hat{\sigma};\mu\nu} - \bar{\theta}_{x;\mu\nu}).
\end{equation}

Using relations (\ref{angles}) and (\ref{wil_act}), eq. (\ref{com_act})
takes the form
\begin{eqnarray}
S_{UH} &  = & \beta \sum_{x;\mu\nu}(1 - \cos\theta_{x;\mu\nu}) -
 2\kappa\sum_{x}\sum_{\mu} \rho_{x}\rho_{x+\hat{\mu}} \cos(\sigma_{x+\hat{\mu}}-
       \sigma_{x} + \theta_{x,\mu}) \nonumber \\
& & + \lambda \sum_{x}(\rho_{x}^{2}-1)^{2} + \sum_{x} \rho_{x}^{2}.
\end{eqnarray}
The  modified compact U(1)--Higgs model, henceforth
called modified scalar QED,  is defined by the action
$S_{MUH}$ which is just $S_{UH}$ with $S_{W}$ replaced by  $S_{MA}$:
\begin{equation}\label{mod_act}
S_{MUH} = S_{UH} + \gamma \sum_{x;\rho} |M_{x;\rho}|.
\end{equation}
In our simulations, the condition $\gamma = \infty$ is
set in (\ref{mod_act}), ensuring that monopoles are completely suppressed.
Notice that in previous investigations \cite{BMM,hmm1} negative
plaquette--values have been suppressed, too.

\section{Phase Diagram}

\begin{figure}
\vspace{6.0cm}
\includegraphics{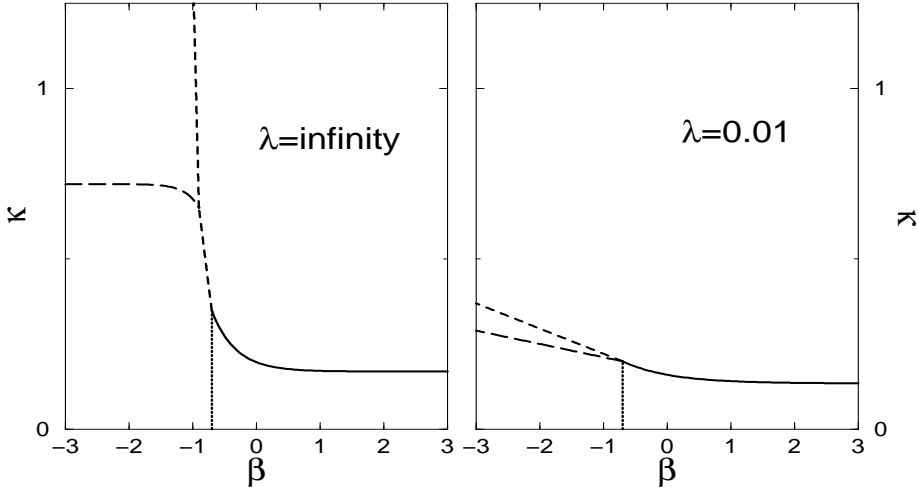}
\caption{The phase diagram for $~\lambda = 0.01~$ and $~\lambda = \infty$.}
\label{fig:phase_1}
\end{figure}

We first explored the phase diagram of the model (\ref{mod_act}) with
$\gamma = \infty$ in the non--linear limit $\lambda = \infty$, where the
radial part $\rho_x$ of the Higgs field is frozen to unity. This was done
with ``hysteresis'' type runs on a $6^4$ lattice. We monitored the average
plaquette, $\langle {\rm Re}( U_{\Box}) \rangle$, and the `average link',
$\langle \phi^{*} U \phi \rangle$.  The phase diagram found in this way is
shown in Figure \ref{fig:phase_1}. At positive $\beta$ we found the phase
transition line separating the Coulomb phase at small $\kappa$ from the
Higgs phase at larger $\kappa$. Because of the complete suppression of
monopoles no confining phase was found, and the Higgs transition line did
not end -- in fact could not end -- somewhere in the interior of the
$(\beta,\kappa)$--plane. This is in contrast to the model without
suppression of monopoles, where the confinement -- Coulomb phase transition
of the pure gauge model turns into a phase transition line that meets the
Higgs transition line, and where the extension of the Higgs transition line
ends in the interior of the $(\beta,\kappa)$--plane. This ending is possible
since before reaching it, the Higgs transition lines separated the
confinement and Higgs phases which are both massive phases, and are
analytically connected. As for the unrestricted model, hysteresis runs do
not give clear evidence for the order of the Higgs transition in the model
with completely suppressed monopoles, though a weak first order transition
seems somewhat more likely.

Though the model is not physical there since the bare gauge coupling
becomes imaginary, one can study the model on the lattice also for negative
$\beta$. In our model with completely suppressed monopoles, nothing special
seems to happen at $\beta=0$. The Higgs transition continues to negative
$\beta$ until $\beta \approx -0.7$. The pure gauge model has a transition
at $\beta^* \approx -0.7$ into a frustrated phase. The Wilson pure gauge action
has a symmetry $\beta \to -\beta$ (with $U_{\Box} \to -U_{\Box}$), but the
monopole term prefers positive $U_{\Box}$, destroying the symmetry and
leading to frustration. This first order transition continues into the
$(\beta,\kappa)$--plane -- we have seen long lived coexisting states at
$(\beta,\kappa) = (-0.705,0.15)$ --, meets the Higgs transition and
continues upward. Around $(\beta,\kappa) = (-0.9,0.65)$ the transition line
splits into two, with one branch soon running almost horizontally, parallel
to the $\beta$--axis. We did not spend any resources to either locate these
transitions precisely or study the properties of the different phases in
the negative $\beta$ half--plane, since we believe that these phases are
irrelevant for the continuum limit. We note, though, that both the Higgs and
Coulomb phases appear to continue into the negative $\beta$ region.

We next explored the phase diagram for a weaker scalar self--coupling,
$\lambda = 0.01$. For the model without the suppression of monopoles it is
known that the phase diagram changes from large to small self--coupling
qualitatively \cite{mitrju}.
While for large $\lambda$ the Higgs transition ends in the
interior of the $(\beta,\kappa)$--plane, for smaller couplings it extends
to, and beyond, the $\beta=0$ axis. The phase diagram for our model with
complete suppression of monopoles is shown in Figure \ref{fig:phase_1}. It
resembles the phase diagram for the non--linear case ($\lambda =\infty$):
the Higgs transition separating Coulomb and Higgs phase at positive
$\beta$, now clearly of first order, continues to negative $\beta$ until it
is met by the  ``frustration'' transition line emanating from
$\beta \approx -0.7$
on the pure gauge axis. Unlike the non--linear case, for $\lambda=0.01$ two
distinct phase transition lines appear to be emerging from this point and
to continue at different angles to larger $\kappa$ and more negative
$\beta$.

In the remainder of this paper we will concentrate on the physics around
the Higgs phase transition, and we will present evidence for the Coulomb
and Higgs character of the phases.
Our results are based on between 4000 and 8000 essentially statistically
independent measurements, after thermalization, for each $\lambda$, $\beta$,
$\kappa$ triplet, except for the points at $\lambda=3.0$, $\beta=-0.5$ where
we have only 2000 measurements.

\section{Potential, Screening Energy and Order Parameter}

In gauge theories, unlike in spin models or scalar field theories, there do
not exist local order parameters to distinguish the different phases.
Rather one has to study non--local objects for this purpose. For pure gauge
theories, for example, area or perimeter behavior of Wilson loops
distinguishes between confining and non--confining phases. However,
it has been known for some time that the Wilson loop criterion
fails in gauge theories where dynamical matter fields screen the potential
\cite{ks,FS}. In such theories, gauge--invariant objects involving
external sources seem to serve better to study the structure of the model.

Therefore, besides computing the usual Wilson loops we also studied
the gauge--invariant two--point function $G(T,R)$, defined by
\begin{equation}\label{eq:GI2P}
G(T,R) = \langle \phi_{x}^{*} \prod_{l \in \Gamma} U_{l}
                 \phi_{y} \rangle , \quad |x - y| = T
\end{equation}
where the $U_{l}$'s are link variables on an oriented path
$\Gamma$ connecting the points $x$ and $y$ as
follows:

\unitlength0.5cm
\begin{picture}(14,4)
\put(7,2){$\Gamma = $}
\put(10,1){\line(1,0){2}}
\put(10,3){\line(1,0){2}}
\put(12,1){\line(0,1){2}}
\put(9.75,0.8){$\times$}
\put(9.75,2.85){$\times$}
\put(9,0.9){$x$}
\put(9,2.9){$y$}
\put(11,0.25){$R$}
\put(12.5,2){$T$}
\end{picture}

\noindent Notice the ordering of arguments in $G(T,R)$. The first gives the
direction in which one side does not contain a string of $U$--fields. We
also computed the gauge--invariant two--point function $G(R,T)$ where the
matter fields are in the same time--slice.

These gauge--invariant functions turn out to be much better  objects to
study confinement and/or Higgs character \cite{EGJJKN} in compact scalar
QED. In our case, they turn out to be very good order parameters for
signalling the Higgs transition.

To possibly improve the signal to noise ratio we not only measured Wilson
loops and the gauge--invariant two--point functions with the ordinary link
fields but also with ``APE--smeared'' links \cite{APE_smear} for the
spatial segments. Use of such smeared links has been extremely helpful in
the study of the heavy quark potential in lattice QCD.

\subsection{Potential}

To be able to study the static potential we have computed on--axis and some
off--axis Wilson loops, the latter only for $\lambda=0.01$. We considered
the distances $R=n$, and in the case of having off--axis loops also
$\sqrt{2} n$ and $\sqrt{5} n$. To try to get a better overlap with the
ground--state in
\eq
 W(\vec R, T) = c \exp\{ -T \cdot V(\vec R) \} + \mbox{subleading terms}
\en
we computed the Wilson loops also with smeared spatial links. We shall
refer to these Wilson loops as ``smeared'' loops. For $\lambda = 3.0$ and
for $\lambda=0.01$ at $\beta=2.5$, actually only smeared loops were
computed.
In simulations of lattice QCD
the use of smeared loops leads to an early plateau in the finite
$T$ approximants (effective potentials)
\eq
V_T(\vec R) = \log \left( \frac{W(\vec R, T)}{W(\vec R, T+1)} \right)
\label{eq:eff_pot}
\en
to the potential $V(\vec R)$  and is
invaluable in the extraction of the potential.

For unsmeared planar Wilson loops, using the fact that they are symmetric
under the exchange of space and time direction, we can estimate how fast
the finite $T$ approximants approach the true potential. For a Yukawa
potential we find
\eq
V_T(R) = V(R) + c R \left[ \frac{1}{T} {\rm e}^{-\mu T} -
\frac{1}{T+1} {\rm e}^{-\mu (T+1)} \right]
\label{eq:T_infty}
\en
{\it i.e.,} an exponential approach. For a Coulomb potential, on the other
hand, the $\mu \to 0$ limit of (\ref{eq:T_infty}), the approach is only
power--like, like $1/T^2$.

For the U(1)--Higgs model studied by us, smeared Wilson loops were of some
use, though not nearly as dramatic as in QCD. They lead to a slightly
earlier plateau in the Higgs phase, and to smaller errors in both Higgs and
Coulomb phases. In the Coulomb phase, the smeared loops gave a somewhat
smaller dependence of $V_T(\vec R)$ on $T$, but they still seemed to
approach an asymptotic value only as $1/T^2$, just like the potential
approximants obtained from normal Wilson loops. We generally did get a
good signal even at the largest distances $R$, out to $T=7$, on the $8^3
\times 16$ lattices that we simulated.

For pure gauge U(1) the perturbative expansion of the lattice potential
leads to
\eq
V^{\rm pert}_T(\vec R) = g^2 V^{\rm Coul}_T(\vec R) \left[ 1 +
\frac{g^2}{4} \right] + {\cal O}(g^6) ,
\label{eq:PT_pot}
\en
where the finite $T$ approximants are defined analogous to
(\ref{eq:eff_pot}), $g^2 = 1/\beta$ and $V^{\rm Coul}_T(\vec R)$ is the
tree--level -- T dependent -- contribution with the coupling
factored out. Possible non--Coulomb
terms appear only at higher orders, and to order $g^4$ we just find a
renormalization of the bare coupling.

Scalar fields, as long as we stay in the Coulomb phase of course, induce
some non--Coulomb terms already at order $g^4$. They are, however, except
possibly very close to the Higgs transition, expected to be small. Since,
as mentioned before, the finite $T$ approximants to the potential approach
their asymptotic value, the true potential, only as $1/T^2$, we decided,
rather than trying to extract this `true' potential and then attempting
fits to it, to make fits of the finite $T$ approximants to the lowest order
perturbative from, but with a renormalized coupling as the (only) fit
parameter,
\eq
V^{\rm MC}_T(\vec R) = g^2_R V^{\rm Coul}_T(\vec R)  ~.
\label{eq:pot_fit}
\en

\begin{table}
\begin{center}
\begin{tabular}{||r|c|c|c|c||} \hline
 $\beta$ & $\kappa$ & $n_{smear}$ & $g^2_R$ & $\chi^2$/ndf \\ \hline \hline
 -0.5  & 0.1850  & 7 & 7.84(3)    & 1.083 \\ \hline
  0.5  & 0.1450  & 0 & 2.892( 7)  & 0.647 \\ \hline
  0.5  & 0.1450  & 5 & 2.896( 2)  & 0.459 \\ \hline
  0.5  & 0.1475  & 0 & 2.887( 6)  & 0.460 \\ \hline
  0.5  & 0.1475  & 5 & 2.889( 4)  & 0.453 \\ \hline
  0.5  & 0.1500  & 0 & 2.867(10)  & 0.455 \\ \hline
  0.5  & 0.1500  & 7 & 2.869( 6)  & 0.693 \\ \hline
  2.5  & 0.1275  & 7 & 0.4490(4)  & 0.639 \\ \hline
  2.5  & 0.1300  & 7 & 0.4489(3)  & 0.239 \\ \hline \hline
\end{tabular}
\end{center}
\caption{The couplings $g^2_R$ in the Coulomb phase from fits as in eq.
         (\protect\ref{eq:pot_fit}) of the finite $T$ approximants to the
         potential to the leading finite volume lattice perturbative values
         for $\lambda=0.01$. Notice that the perturbative value for
         $\beta=2.5$ is 0.44 plus ${\cal O}(0.4^3)$.}
\label{tab:g_ren1}
\medskip\noindent
\end{table}

\begin{table}
\begin{center}
\begin{tabular}{||r|c|c|c|c||} \hline
 $\beta$ & $\kappa$ & $n_{smear}$ & $g^2_R$ & $\chi^2$/ndf \\ \hline \hline
 -0.5  & 0.3100  & 7 & 7.81(3)    & 0.384 \\ \hline
 -0.5  & 0.3150  & 7 & 7.72(6)    & 0.979 \\ \hline
 -0.5  & 0.3175  & 7 & 7.68(7)    & 0.921 \\ \hline
 -0.5  & 0.3200  & 7 & 7.59(5)    & 0.502 \\ \hline
  0.5  & 0.2100  & 7 & 2.885( 5)  & 0.432 \\ \hline
  0.5  & 0.2125  & 7 & 2.873( 6)  & 0.466 \\ \hline
  2.5  & 0.1750  & 7 & 0.4486(3)  & 0.241 \\ \hline
  2.5  & 0.1775  & 7 & 0.4489(3)  & 0.490 \\ \hline
  2.5  & 0.1800  & 7 & 0.4469(4)  & 0.553 \\ \hline \hline
\end{tabular}
\end{center}
\caption{The couplings $g^2_R$ in the Coulomb phase from fits as in eq.
         (\protect\ref{eq:pot_fit}) of the finite $T$ approximants to the
         potential to the leading finite volume lattice perturbative values
         for $\lambda=3.0$.
         Notice that the perturbative value for $\beta=2.5$ is 0.44
         plus ${\cal O}(0.4^3)$.}
\label{tab:g_ren2}
\medskip\noindent
\end{table}

\begin{figure}
\vspace{7.0cm}
\includegraphics{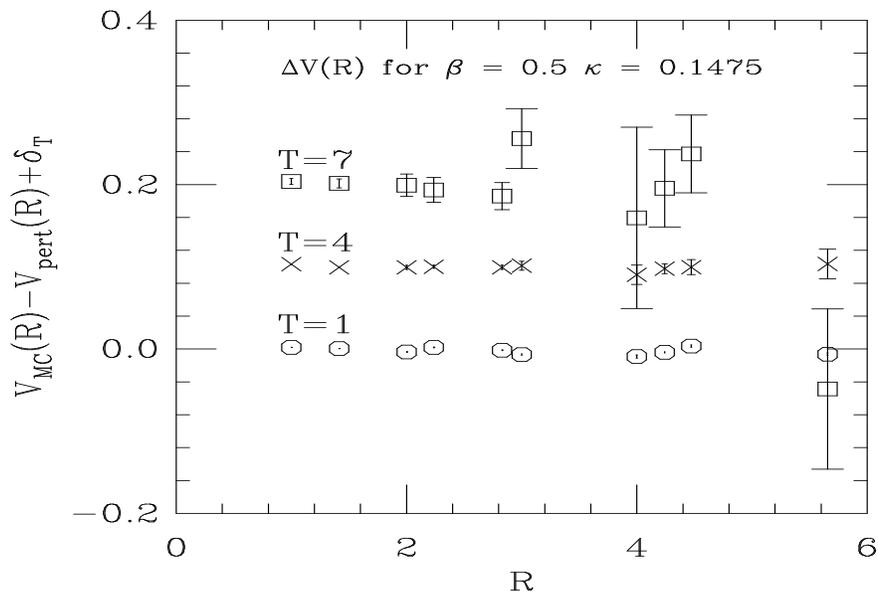}
\caption{The difference between the measured and fitted effective
         potentials for $T=1$, $4$ and $7$, for $\beta=0.5$, $\lambda=0.01$
         and $\kappa=0.1475$. The values for $T=4$ and $7$ have been
         displaced by $\delta_T=0.1$ and 0.2 vertically for better viewing.}
\label{fig:del_V}
\end{figure}

To fit also the effective potential from smeared Wilson loops, we computed
the `smeared' analogue of $V^{\rm Coul}_T(\vec R)$ with the smeared links,
constructed as in the Monte Carlo simulations, but expanded to lowest
order in perturbation theory. These fits work very well (for positive
$\beta$) once we exclude the approximants from $T=1$ and the shortest
distance, $R=1$. The results are listed in Table \ref{tab:g_ren1} for
$\lambda=0.01$ and in Table \ref{tab:g_ren2} for $\lambda=3.0$. To illustrate
the quality of the fits we show in Figure \ref{fig:del_V} the difference
between the measured and fitted effective potentials for $T=1$, which was
{\it not} included in the fit, and for $T=4$ and $7$, for $\beta=0.5$,
$\lambda=0.01$ and $\kappa=0.1475$. For $\beta=2.5$, the bare coupling is
$g^2=0.4$ and we expect perturbation theory to be applicable. Indeed, the
difference between the values obtained from the fits in Tables
\ref{tab:g_ren1} and \ref{tab:g_ren2} to the perturbative prediction of
0.44 from (\ref{eq:PT_pot}) is certainly compatible with coming from higher
order perturbative corrections.

As can be seen from Table \ref{tab:g_ren1} the results from normal and
smeared loops agree within errors with the smeared loops giving somewhat
smaller errors. The dependence on $\kappa$, {\it i.e.,} the scalar fields,
is rather weak with $g^2_R$ decreasing slightly as the scalars become
lighter (as the Higgs transition is approached) as one would expect.

In Table \ref{tab:g_ren1} no $g^2_R$ is quoted for normal Wilson loops for
$\beta=-0.5$. The reason is that several loops, including the plaquette,
were negative there. In fact, exactly those of the small loops were
negative, that would become positive under the transformation that makes the
symmetry $\beta \to -\beta$ for the Wilson model, {\it i.e.}, without the
suppression of monopoles. Smearing got rid of this problem. However, the
Wilson loops remained rather noisy compared to those obtained at positive
gauge coupling and the approach to an asymptotic value of the
effective potentials (\ref{eq:eff_pot}) was not monotonic. This fact points
to a non--positive transfer matrix, as one might expect for the imaginary
bare gauge coupling one has for negative $\beta$. A similar, non--monotonic
behavior was observed in the Higgs phase, as will be discussed further below.
We should add here, that contrary to all other cases, only time separations
up to $T=4$ could be included in the fit for $g^2_R$. For larger $T$ we
did not get a signal for the effective potential except for the smallest
distances.

In the Higgs phase, the potential is a screened, or Yukawa potential.
There we made fits to finite $T$ approximants of the lattice Yukawa potential
\eq
V^{\rm MC}_T(\vec R) = g^2_R \cdot V^{\rm Yuk}_T(\vec R)  ~,
\label{eq:Yuk_fit}
\en
\noindent obtained from lowest order Wilson loops with the vector boson
propagator in $R_{\xi}$ gauge with $\xi=1$ given by
\eq
G_{\mu\nu}(p) = \frac{\delta_{\mu \nu}}{4 \sum_\mu \sin^2 \frac{p_{\mu}}{2}
+ 4\sinh^2 \frac{\mu_{\gamma}}{2} },
\label{eq:mass_prop}
\en
\noindent with $\mu_{\gamma}$ the mass of the gauge boson.
We used as a fitting parameter
$m_{\gamma} = 2\sinh \frac{\mu_{\gamma}}{2}$.

\begin{table}
\begin{center}
\begin{tabular}{||r|c|c|c|c|c||} \hline
 $\beta$ & $\kappa$ & $n_{smear}$ & $g^2_R$ & $m_{\gamma}$ &
 $\chi^2$/ndf \\ \hline \hline
  0.5  & 0.1506  & 0 & 2.77(5)   & 0.92(4)   & 1.009 \\ \hline
  0.5  & 0.1506  & 0 & 3.00(3)   & 1.12(2)   & 1.309 \\ \hline
  0.5  & 0.1506  & 7 & 2.93(1)   & 1.06(1)   & 1.112 \\ \hline
  0.5  & 0.1506  & 7 & 2.88(2)   & 1.02(2)   & 2.231 \\ \hline
  0.5  & 0.1525  & 0 & 3.00(1)   & 1.71(1)   & 0.461 \\ \hline
  0.5  & 0.1525  & 0 & 2.96(3)   & 1.69(3)   & 0.605 \\ \hline
  0.5  & 0.1525  & 7 & 2.89(2)   & 1.63(2)   & 0.499 \\ \hline
  0.5  & 0.1525  & 7 & 2.78(4)   & 1.55(3)   & 1.159 \\ \hline
  0.5  & 0.1550  & 0 & 3.00(3)   & 2.17(2)   & 0.735 \\ \hline
  0.5  & 0.1550  & 7 & 2.94(1)   & 2.13(1)   & 0.888 \\ \hline
  0.5  & 0.1550  & 7 & 2.94(2)   & 2.12(2)   & 2.256 \\ \hline
  0.5  & 0.1575  & 0 & 2.85(2)   & 2.42(2)   & 0.679 \\ \hline
  0.5  & 0.1575  & 0 & 2.84(4)   & 2.40(3)   & 1.170 \\ \hline
  0.5  & 0.1575  & 7 & 2.85(1)   & 2.42(1)   & 0.705 \\ \hline
  0.5  & 0.1575  & 7 & 2.85(1)   & 2.41(1)   & 1.396 \\ \hline
  2.5  & 0.1350  & 7 & 0.447(1)  & 0.415(8)  & 1.313 \\ \hline
  2.5  & 0.1375  & 7 & 0.445(1)  & 0.550(3)  & 0.315 \\ \hline
  2.5  & 0.1375  & 7 & 0.445(1)  & 0.550(3)  & 0.839
\\ \hline \hline
\end{tabular}
\end{center}
\caption{The fit parameters from fits to the finite $T$ approximants
         to the lattice Yukawa potential, obtained from lowest order
         Wilson loops with a massive gauge boson propagator, for
         $\lambda=0.01$. The first line for each data set is the
         fit with smallest $\chi^2$/ndf, and the second line comes
         from $T=7$, unless the fits are identical.}
\label{tab:Yuk_fit_2b}
\medskip\noindent
\end{table}

\begin{figure}
\vspace{7.0cm}
\includegraphics{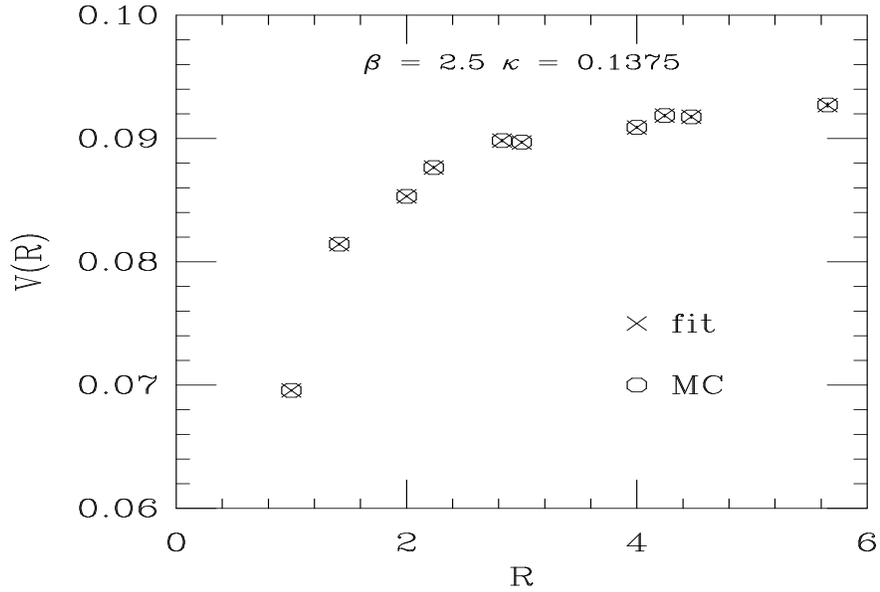}
\caption{Monte Carlo potential from smeared Wilson loops and the fit to
         the effective potential extracted from lowest order perturbative
         Wilson loops with a massive propagator, for $\beta=2.5$,
         $\lambda=0.01$ and $\kappa=0.1375$.}
\label{fig:Yuk_fit_2}
\end{figure}

These fits, tabulated in Table \ref{tab:Yuk_fit_2b} and
\ref{tab:Yuk_fit_3b}, worked  well for the $\lambda=0.01$, $\beta=0.5$ data
where $m_\gamma$ is relatively large and a plateau is achieved quite soon,
but they worked much better for the data at the larger $\beta$ value,
$\beta=2.5$. A sample fit is shown in Figure \ref{fig:Yuk_fit_2}.  Notice
the violations of rotational symmetry, especially at short distances, which
are well followed by the lattice Yukawa potential. A fit to the continuum
Yukawa potential $\frac{1}{R} {\rm e}^{-m_{\gamma} R}$ would not have
worked. For data in the Coulomb phase, the Yukawa fits, correctly gave a
vanishing photon mass, within error.

\begin{table}
\begin{center}
\begin{tabular}{||r|c|c|c|c|c||} \hline
 $\beta$ & $\kappa$ & $n_{smear}$ & $g^2_R$ & $m_{\gamma}$ &
 $\chi^2$/ndf \\ \hline \hline
  0.5  & 0.2150  & 4 & 2.98( 7)  & 0.43(13)   & 0.823 \\ \hline
  0.5  & 0.2175  & 4 & 2.94( 6)  & 0.42(10)   & 0.048 \\ \hline
  0.5  & 0.2200  & 4 & 2.93( 2)  & 0.40( 3)   & 0.671 \\ \hline
  0.5  & 0.2225  & 7 & 2.93( 2)  & 0.44( 3)   & 0.321 \\ \hline
  0.5  & 0.2250  & 4 & 2.93( 1)  & 0.46( 2)   & 0.031 \\ \hline
  2.5  & 0.1825  & 7 & 0.449(1)  & 0.119( 8)  & 0.478 \\ \hline
  2.5  & 0.1850  & 7 & 0.448(1)  & 0.154(14)  & 0.031 \\ \hline
  2.5  & 0.1875  & 7 & 0.448(1)  & 0.179(16)  & 0.790
\\ \hline \hline
\end{tabular}
\end{center}
\caption{The fit parameters from fits to the finite $T$ approximants
         to the lattice Yukawa potential, obtained from lowest order
         Wilson loops with a massive gauge boson propagator, for
         $\lambda=3.0$.}
\label{tab:Yuk_fit_3b}
\medskip\noindent
\end{table}

It maybe worthwhile to note that the attempt to fit to the `naive' lattice
Yukawa potential, the $T \to \infty$ limit of $V^{\rm Yuk}_T(\vec R)$,
\eq
V_{naive}(\vec R) = V_0 +  g^2_R \frac{1}{L^3} \sum_{\vec p} \frac{1 -
 \cos( \vec p \vec R)}{4 \sum_{1 \leq k \leq 3} \sin^2 \frac{p_k}{2}
 + m^2_{\gamma}}
\label{eq:Yuk_fit_naive}
\en
was successful only for $\lambda=0.01$ and $\beta=0.5$. For $\beta=2.5$ and
both values of $\lambda$, as well as for $\beta=0.5$ and $\lambda=3.0$ we
could not obtain fits with a decent $\chi^2$ using
eq.~(\ref{eq:Yuk_fit_naive}). This observation is in disagreement with some
statements in the literature \cite{JJLNV}. A possible explanation might be
that smaller statistical errors in our case lead to a better
discrimination between different fit formulae, thus excluding the `naive'
fits.

\begin{figure}
\vspace{7.0cm}
\includegraphics{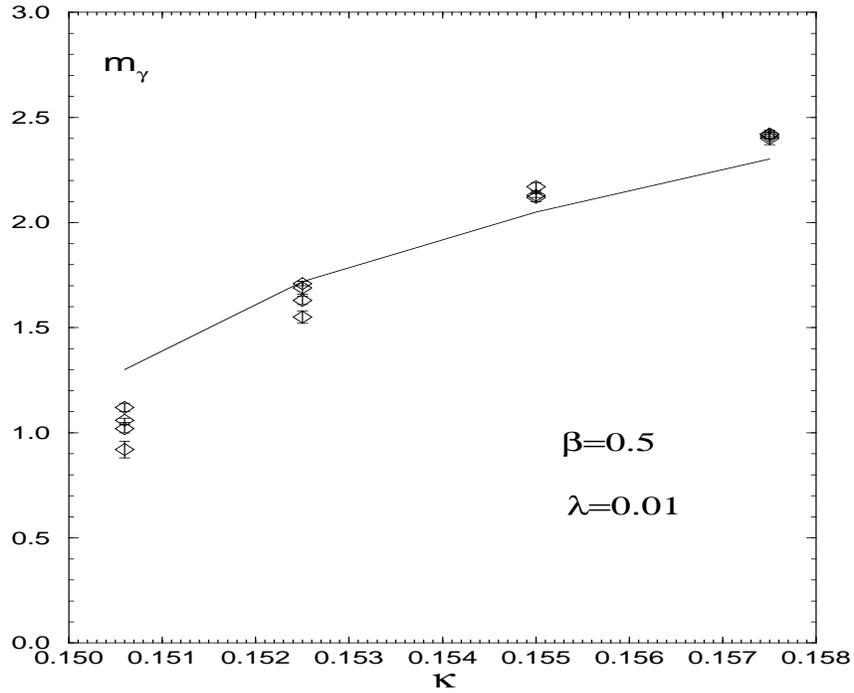}
\vspace{-0.5cm}
\caption{
The parameter $m_{\gamma}$ in the Higgs phase. Diamonds correspond
to $m_{\gamma}$ values taken from Table \protect\ref{tab:Yuk_fit_2b}.
The solid line corresponds to r.h.s. of
eq.~(\protect\ref{m_gamma}).
}
\label{fig:m_gamma}
\end{figure}

In the quasiclassical perturbative approach one expects
\eq
m^2_{\gamma} \simeq \frac{2\kappa}{\beta}\cdot \langle \phi^{\ast}\phi \rangle.
\label{m_gamma}
\en
Perturbative arguments suggest that this equality should be valid  in the
weak coupling region when the dimensionless mass $~\mu_{\gamma}~$ is
sufficiently small, {\it i.e.}, $~m_{\gamma} \simeq \mu_{\gamma}$.  It is
interesting to note that this equality seems to hold even in the region
where $~m_{\gamma}~$ is not very small. In Figure \ref{fig:m_gamma} we
compare the values of $~m_{\gamma}~$ taken from Table \ref{tab:Yuk_fit_2b}
with that calculated via eq.~(\ref{m_gamma}) at $~\beta =0.5~$ and $~\lambda
=0.01$. Qualitatively, the agreement is quite reasonable, indicating
consistency of the quasiclassical estimation.

\begin{figure}
\vspace{7.0cm}
\includegraphics{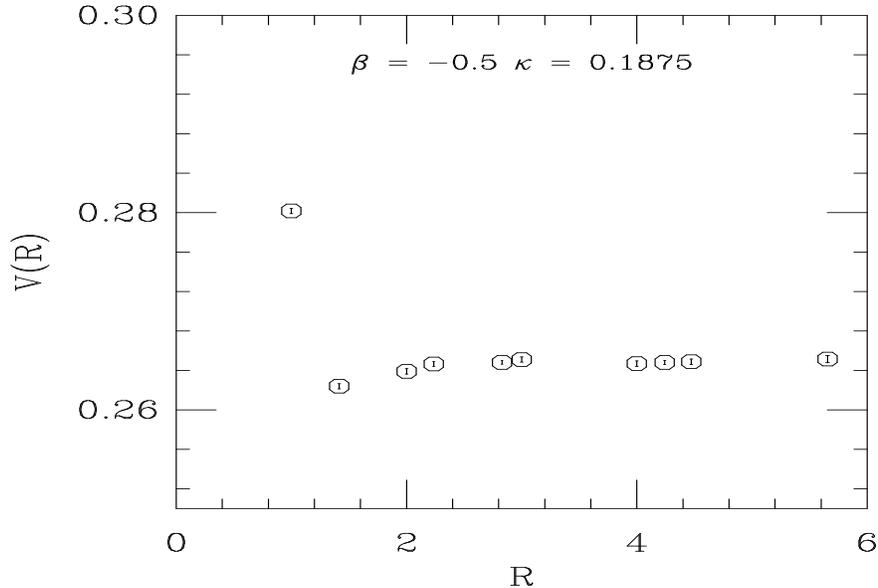}
\caption{Monte Carlo potential from smeared Wilson loops for $\beta=-0.5$,
         $\lambda=0.01$ and $\kappa=0.1875$.}
\label{fig:Yuk_fit_3}
\end{figure}

For $\beta=-0.5$ the potentials looked rather strange, with the point at
$R=1$ being the highest. An example is shown in Figure \ref{fig:Yuk_fit_3}.
The potentials from smeared and normal Wilson loops, where available,
agreed. As in the Coulomb phase, the potential approximants show a
non--monotonic behavior with $T$, though a plateau is reached rather
quickly. Even ignoring the $R=1$ point, no good fits could be obtained, and
hence no conclusions drawn.

\subsection{Screening Energy}

The gauge invariant two--point function (\ref{eq:GI2P}) can be interpreted
as an external static charge propagating forward in (Euclidean) time. The
presence of the external charge manifests  itself as an energy increase
$\mu$ of the lowest state of the system. The screening of an external
source may be studied from the exponential fall--off of $G(T,R)$ for fixed
$R$ and for $T$ large enough:
\begin{equation}\label{eq:mu}
G(T,R) \sim f(R) \exp(-\mu T).
\end{equation}
For smaller $T$ there are subleading exponentials to this. $\mu$ is the
lowest energy in the fields that screen an external charge, and is
therefore called the screening energy.

In a confining phase in the presence of matter fields or in the Higgs phase,
the external charges are always completely screened and one therefore finds
that
\eq\label{eq:mu_V}
\mu = \frac{1}{2} V(\infty)
\en
with $V(R)$ the static potential that can be extracted from Wilson loops.
In the Coulomb phase the screening energy is the energy of a bound--state of
a light charge, represented by the field $\phi^{*}$ in (\ref{eq:GI2P})
and the static charge, corresponding to the straight time--like segment.
With $m_c$ the mass of the light charged particle the screening energy is
then
\eq
\mu = \frac{1}{2} V(\infty) + m_c - E_b
\en
with $E_b$ the binding energy. One therefore expects $\mu > V(\infty)/2$ in
the Coulomb phase.

\begin{table}
\centering
\begin{tabular}{||r|c|c|c|c|c||}\hline
 $\beta,~\kappa \setminus \mu $
          & $R=1$ & $R=2$ & $R=3$ & $R=4$ & $\frac{1}{2}V(R=4)$ \\ \hline
\hline
 -0.5,~0.1850 & 0.1635(1) & 0.1634(1) & 0.1632(3) & 0.1630(3) & 0.1637(10)
\\ \hline
 -0.5,~0.1875 & 0.1325(1) & 0.1327(1) & 0.1330(2) & 0.1329(2) & 0.1322(5)
\\ \hline
 -0.5,~0.1900 & 0.1123(1) & 0.1128(2) & 0.1128(2) & 0.1129(3) & 0.1117(3)
\\ \hline
  0.5,~0.1450 & 1.25(16)  & 0.94(25)  & 1.37(32)  & 1.60(10)  & 0.34(1)
\\ \hline
  0.5,~0.1475 & 1.00(5)   & 0.76(7)   & 0.68(35)  & 0.57(49)  & 0.32(1)
\\ \hline
  0.5,~0.1500 & 0.63(5)   & 0.65(5)   & 0.59(10)  & 0.47(10)  & 0.33(1)
\\ \hline
  0.5,~0.1506 & 0.2437(5) & 0.2442(6) & 0.2447(8) & 0.2412(10)& 0.242(3)
\\ \hline
  0.5,~0.1525 & 0.1844(2) & 0.1841(2) & 0.1842(3) & 0.1842(4) & 0.183(1)
\\ \hline
  0.5,~0.1550 & 0.1484(1) & 0.1486(1) & 0.1487(2) & 0.1485(3) & 0.1487(6)
\\ \hline
  0.5,~0.1575 & 0.1264(1) & 0.1265(1) & 0.1264(1) & 0.1263(1) & 0.1267(3)
\\ \hline \hline
\end{tabular}
\caption{
         Screening energy $\mu$ for $\lambda=0.01$ extracted
         from fits with unsmeared operators for
         different $R$ values and $\kappa$, generally with $T \ge 4$.
}
\label{tab:scrn_1}
\end{table}

\begin{table}
\centering
\begin{tabular}{||r|c|c|c|c|c||}\hline
 $\beta,~\kappa \setminus \mu $
          & $R=1$ & $R=2$ & $R=3$ & $R=4$ & $\frac{1}{2}V(R=4)$ \\ \hline
\hline
 -0.5,~0.1850 & 0.1635(1) & 0.1637(1) & 0.1637(1) & 0.1637(1)& 0.1634(2)
\\ \hline
 -0.5,~0.1875 & 0.1324(1) & 0.1325(1) & 0.1326(1) & 0.1326(1) & 0.1323(1)
\\\hline
 -0.5,~0.1900 & 0.1124(1) & 0.1124(1) & 0.1122(1) & 0.1124(1) & 0.1123(1)
\\ \hline
  0.5,~0.1450 & 1.31(13)  &  0.73(12) & 1.49(7)   & 1.37(7)   & 0.332(1)
\\ \hline
  0.5,~0.1475 & 1.05(6)   & 0.98(6)   & 1.00(9)   & 1.05(10)  & 0.332(2)
\\ \hline
  0.5,~0.1500 & 0.64(5)   & 0.65(5)   & 0.49(4)   & 0.64(7)   & 0.329(2)
\\ \hline
  0.5,~0.1506 & 0.2435(5) & 0.2438(6) & 0.2438(6) & 0.2430(6) & 0.2426(7)
\\ \hline
  0.5,~0.1525 & 0.1841(2) & 0.1840(3) & 0.1842(3) & 0.1845(2) & 0.1838(3)
\\ \hline
  0.5,~0.1550 & 0.1484(1) & 0.1484(1) & 0.1485(1) & 0.1485(2) & 0.1485(2)
\\ \hline
  0.5,~0.1575 & 0.1263(1) & 0.1264(1) & 0.1264(1) & 0.1264(1) & 0.1263(2)
\\ \hline \hline
\end{tabular}
\caption{
         Screening energy $\mu$ for $\lambda=0.01$ extracted
         from fits with smeared operators for
         different $R$ values and $\kappa$, with $T \ge 4$.
}
\label{tab:scrn_2}
\end{table}

\begin{figure}
\epsffile{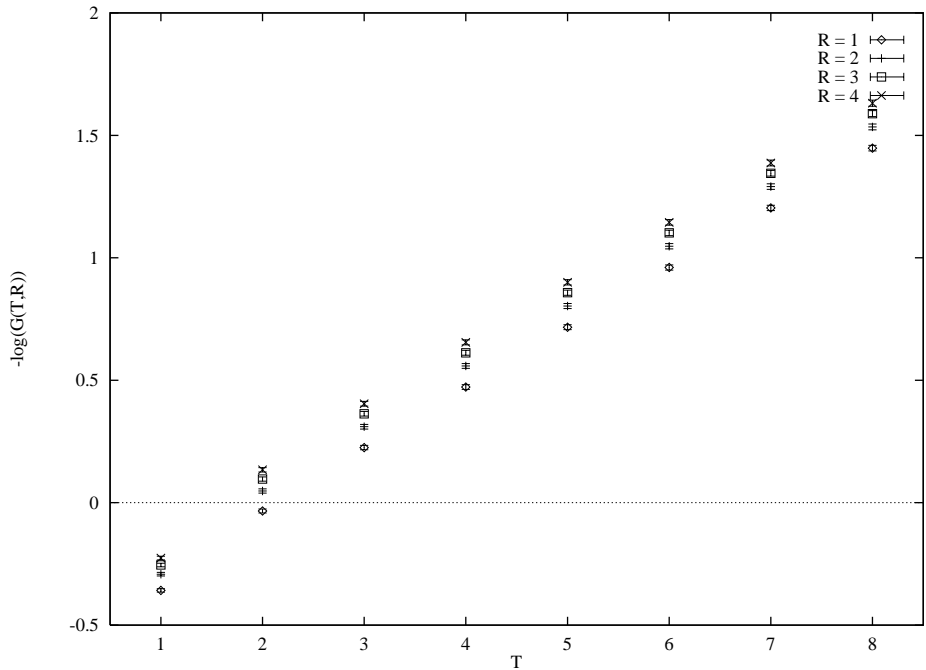}
\caption{The logarithm of $G(T,R)$ plotted as a function of $T$ for
         various $R$ at $\kappa = 0.1506$,
$\beta =0.5$ and $\lambda =0.01$ .}
\label{dist}
\end{figure}

The gauge--invariant two--point function $G(T,R)$ could be calculated quite
accurately and showed in both the free charge phase and the Higgs phase the
expected exponential falloff as a function of $T$, for $T$ large enough
with $R$ fixed as in eq.~(\ref{eq:mu}). Figure~\ref{dist} shows the behavior
of the two--point function as a function of $T$ and different $R$ for
$\kappa = 0.1506$, $\beta=0.5$ and $\lambda=0.01$ near the phase
transition. The screening energy, as extracted from different distances $R$,
is tabulated in Tables \ref{tab:scrn_1} and \ref{tab:scrn_2} for unsmeared
and smeared operators, respectively. The last column in both Tables is an
estimation of the selfenergy of an isolated charge
$E_{q}=\frac{1}{2}V(R=\infty)$ from the potential $V(R=4,T=5)$.
The screening energy as defined by eq.~(\ref{eq:mu}) is supposed
to be independent of the distance $R$ in the gauge invariant two--point
function $G(T,R)$ \cite{EGJJKN}. Our results are in general agreement with
this expectation. In the Coulomb phase, where $\mu$ is rather large and
hence less well determined since $G(T,R)$ becomes noisier at larger $T$,
there are some variations, but clearly no definite trend. We also note that
both smeared and unsmeared operators work equally well leading to
consistent values of $\mu$, except for the `problem points' mentioned
above. We therefore expect that with much better statistics $\mu$ would
again turn out to be independent of $R$.

\begin{figure}
\epsffile{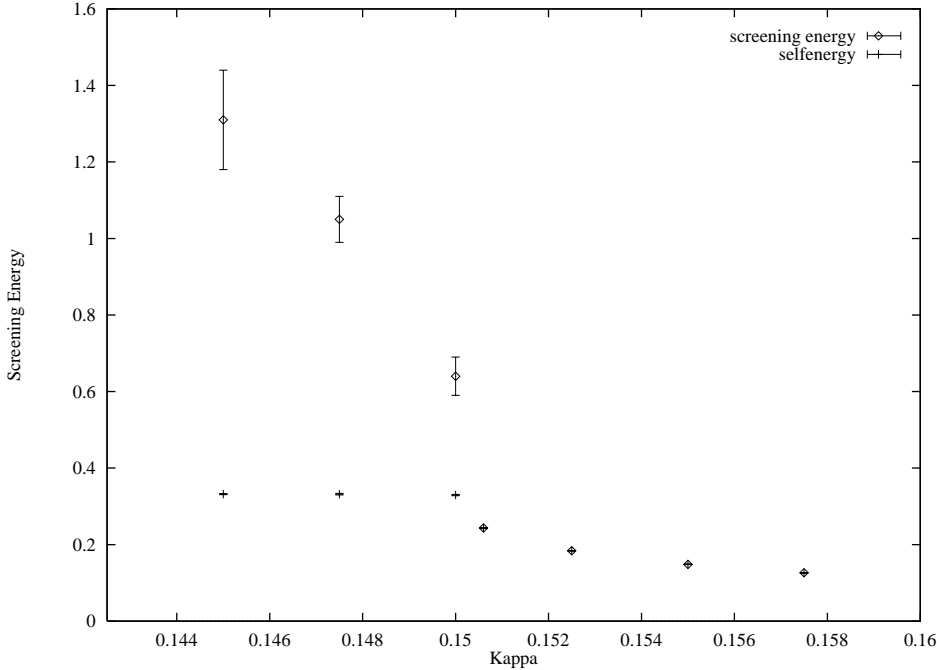}
\caption{The screening energy $\mu$ and the selfenergy of an isolated
         external charge, $E_q$, approximated by $V(R=4)/2$, as a function
         of $\kappa$ for $\beta = 0.5$ and $\lambda=0.01$. $\mu$ is
         determined from the exponential decay of $G(T,R=1)$.}
\label{screen_1}
\end{figure}

In Figure~\ref{screen_1}, we show the screening energy $\mu$ as a function of
$\kappa$, extracted from fits to $\log(G(T,R))$. Also shown in the figure
is $V(R=4)/2$, our approximation to $E_q = V(\infty)/2$, the selfenergy of
an isolated external charge. In the Higgs phase, $V(R=4)/2$ is expected to
be an excellent approximation to $V(\infty)/2$, and as expected $\mu$ and
$V(\infty)/2$ agree very well, (\ref{eq:mu_V}).  In the Coulomb phase,
$V(\infty)/2$ is slightly higher than $V(R=4)/2$ but, as expected, the
screening energy $\mu$ is clearly much larger.

For $\lambda=3.0$ and all $\beta$--values, and for $\lambda=0.01$ at
$\beta=2.5$, due to an oversight we only computed $G(R,T)$ used for the
Fredenhagen--Marcu order parameter, but not $G(T,R)$. Therefore we have no
data for the screening energy for those parameter values.

\subsection{Fredenhagen--Marcu Order Parameter}

Fredenhagen and Marcu found that a suitable criterion to study confinement
in gauge theories with matter fields \cite{FM} can be formulated in terms
of the ratio
\begin{equation}\label{eq:FM}
\rho_{FM}(R,T) = \frac{G(R,T)}{W(R,2T)^{1/2}}.
\end{equation}
This ratio has the asymptotic behavior
\eq
\rho_{FM}(R,T) \conlim \left\{ \begin{array}{ll}
\rho_{FM}^{\infty} ~~~~~ & \mbox{confined or Higgs phase} \\
0 & \mbox{Coulomb phase} \end{array} \right.  ~~~.
\en
The physical interpretation is that one of two charges is removed as
$R\to\infty$. If free charges can exist the overlap of the remaining
one--charge state with the vacuum is zero and the ratio (\ref{eq:FM})
vanishes. If the remaining charge is screened as in a confined or Higgs
phase the ratio attains a finite, non--zero limit. In both cases the
division by the Wilson loop ensures the cancellation of the perimeter law
behavior in $G(R,T)$. The asymptotic value $\rho_{FM}^{\infty}$ is
therefore called the vacuum overlap order parameter. The ratio between $T$
and $R$ is held fixed in the limiting process to ensure projection onto the
state with lowest energy.

\begin{figure}
\epsffile{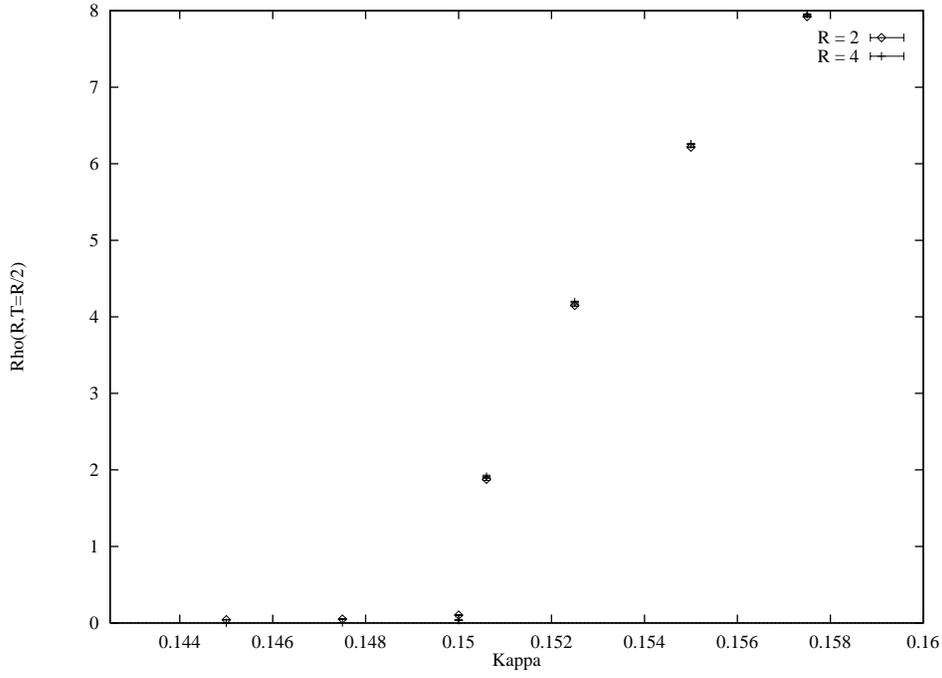}
\caption{The Fredenhagen--Marcu parameter $\rho_{FM}(R,T)$ as a function
         of $\kappa$ for $\beta=0.5$ and $\lambda=0.01$.}
\label{fried_1}
\end{figure}

\begin{figure}
\epsffile{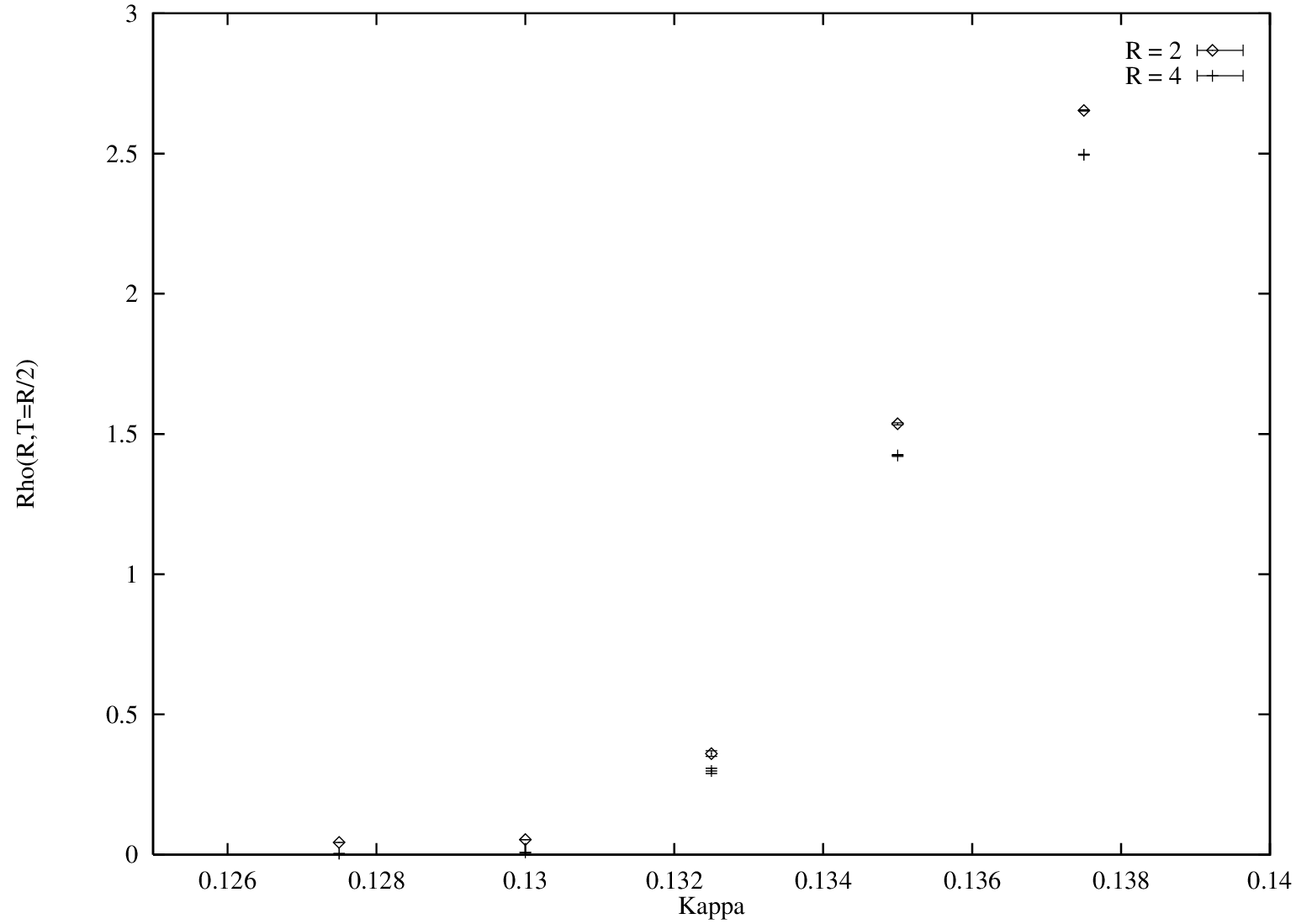}
\caption{The Fredenhagen--Marcu parameter $\rho_{FM}(R,T)$ as a function
         of $\kappa$ for $\beta=2.5$ and $\lambda=0.01$.}
\label{fried_3}
\end{figure}

\begin{figure}
\epsffile{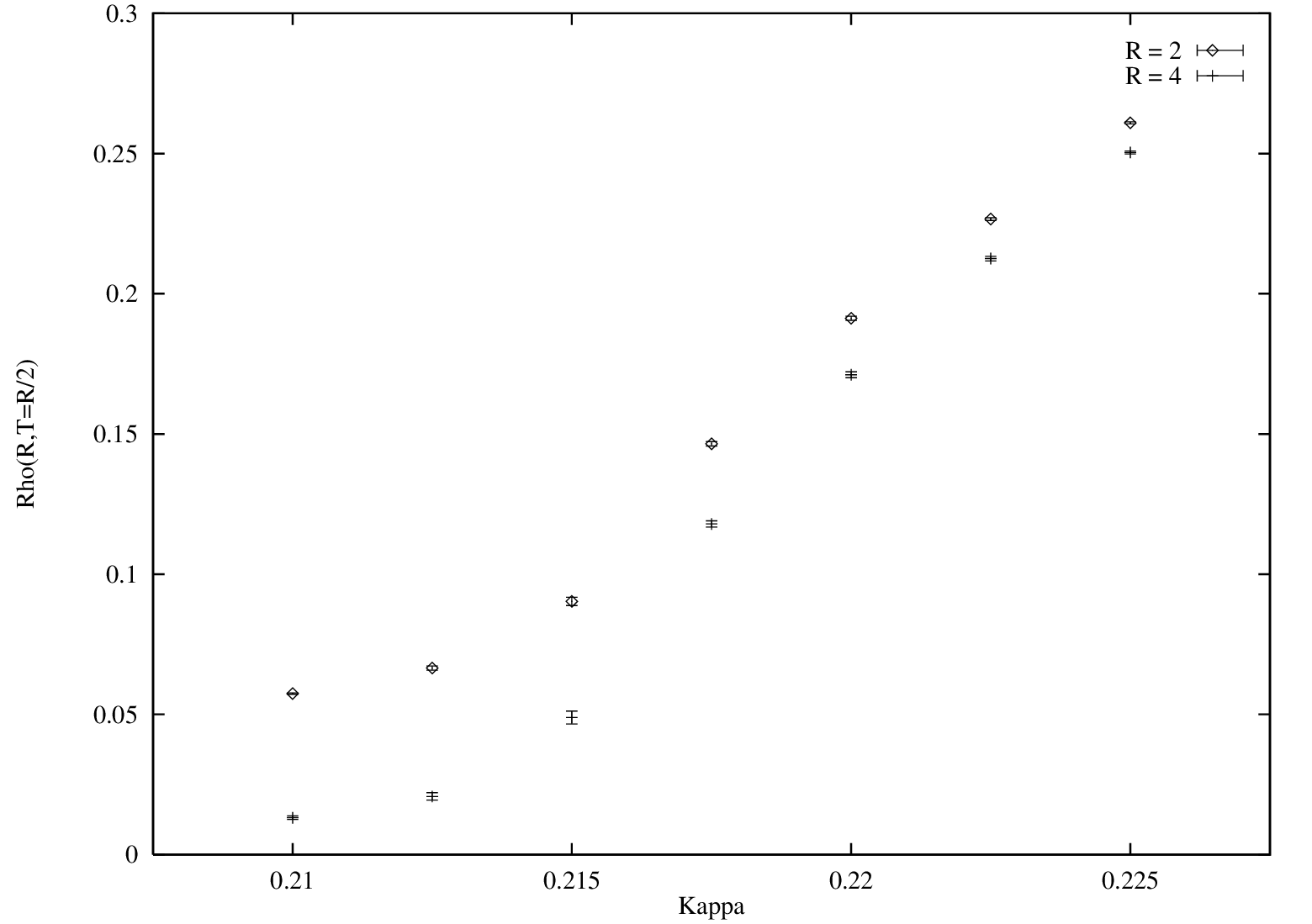}
\caption{The Fredenhagen--Marcu parameter $\rho_{FM}(R,T)$ as a function
         of $\kappa$ for $\beta=0.5$ and $\lambda=3.0$.}
\label{fried_5}
\end{figure}

\begin{figure}
\epsffile{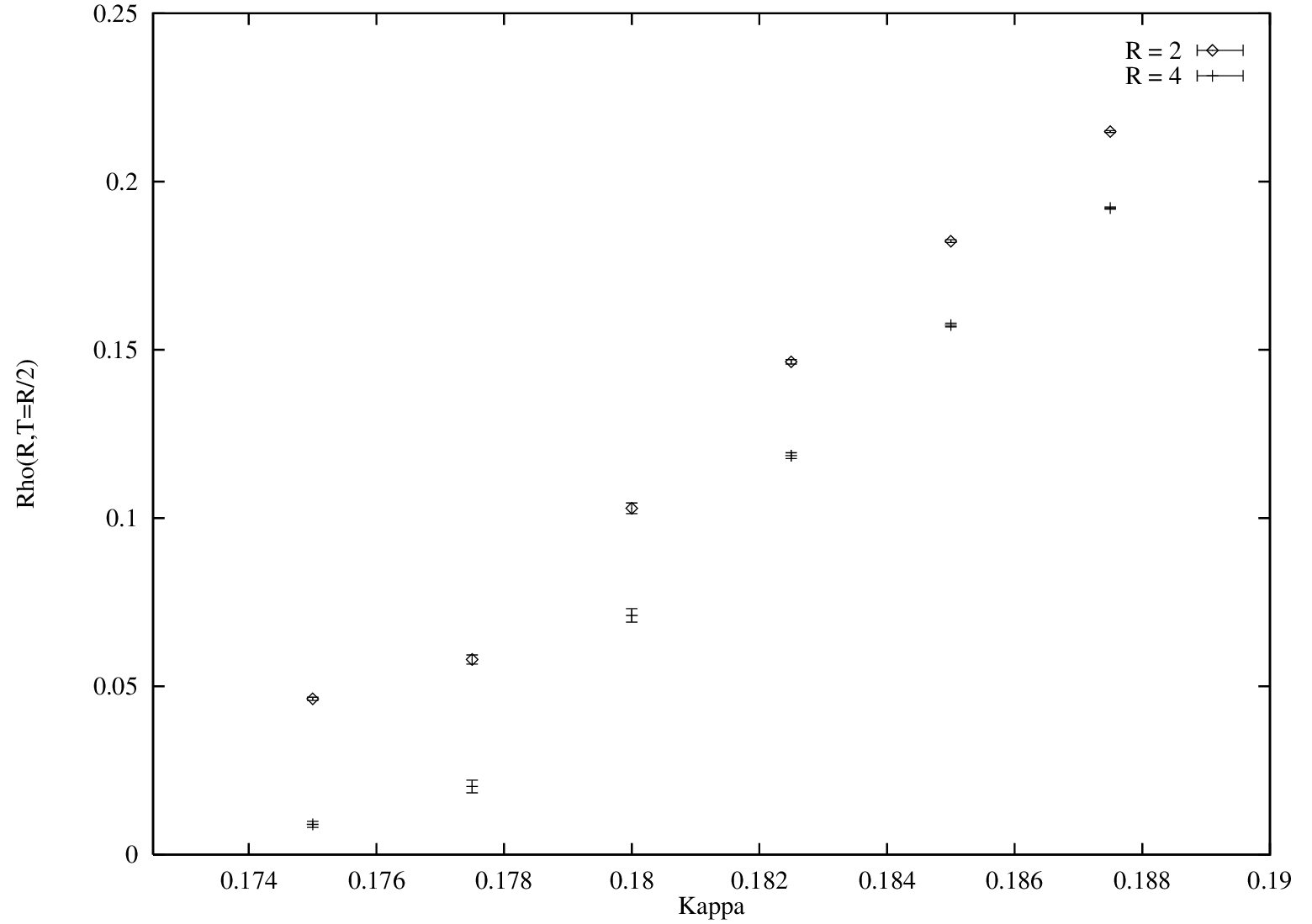}
\caption{The Fredenhagen--Marcu parameter $\rho_{FM}(R,T)$ as a function
         of $\kappa$ for $\beta=2.5$ and $\lambda=3.0$.}
\label{fried_6}
\end{figure}

The Fredenhagen--Marcu parameter $\rho_{FM}(R,T)$ turns out to be a very
good parameter signalling the Higgs phase transition \cite{EGJJKN}. As our
lattices were restricted to $8^{3}\times 16$, we were able to calculate
this only up to $\rho_{FM}(4,2)$. But as Figures~\ref{fried_1} to
\ref{fried_6} indicate, this quantity behaves like an order parameter.

Our data for $\lambda=0.01$ are tabulated in Table \ref{tab:FM_2}
from computations with normal and smeared space--like links
and in Table \ref{tab:FM_3} for $\lambda=3.0$ with smeared space--like links.

\begin{table}
\centering
\begin{tabular}{||r|c|c|c|c||}\hline
 $\beta,~\kappa \setminus \rho_{FM} $
          & $R=2$,norm & $R=4$,norm & $R=2$,smear & $R=4$,smear  \\ \hline
\hline
 -0.5,~0.1850 & 0.278(2)  & 0.007(6)  & 0.0542(4) & 0.0030(8) \\ \hline
 -0.5,~0.1875 & 21.137(5) & 13.886(5) & 15.082(3) & 13.015(3) \\ \hline
 -0.5,~0.1900 & 19.12(4)  & 19.19(6)  & 17.308(3) & 15.279(3) \\ \hline
  0.5,~0.1450 & 0.0654(2) & 0.0024(3) & 0.0420(1) & 0.0025(1) \\ \hline
  0.5,~0.1475 & 0.0786(3) & 0.0041(3) & 0.0510(2) & 0.0045(2) \\ \hline
  0.5,~0.1500 & 0.091(4)  & 0.020(3)  & 0.102(5)  & 0.036(6)  \\ \hline
  0.5,~0.1506 & 1.89(2)   & 1.93(2)   & 1.88(2)   & 1.91(2)   \\ \hline
  0.5,~0.1525 & 4.47(1)   & 2.673(9)  & 4.148(9)  & 4.19(1)   \\ \hline
  0.5,~0.1550 & 5.556(5)  & 4.652(5)  & 6.217(5)  & 6.258(6)  \\ \hline
  0.5,~0.1575 & 6.636(2)  & 5.254(2)  & 7.923(3)  & 7.952(3)  \\ \hline
  2.5, 0.1275 &           &           & 0.0433(2) & 0.0034(3) \\ \hline
  2.5, 0.1300 &           &           & 0.0530(4) & 0.0080(6) \\ \hline
  2.5, 0.1325 &           &           & 0.36(1)   & 0.30(1)   \\ \hline
  2.5, 0.1350 &           &           & 1.536(4)  & 1.424(3)  \\ \hline
  2.5, 0.1375 &           &           & 2.654(2)  & 2.496(2)  \\ \hline \hline
\end{tabular}
\caption{The Fredenhagen--Marcu parameter $\rho_{FM}(R,T)$
         for $\lambda=0.01$ extracted from unsmeared (normal) and
         smeared operators for $R=2$ and $R=4$ with $T=R/2$.}
\label{tab:FM_2}
\end{table}

\begin{table}
\centering
\begin{tabular}{||r|c|c||}\hline
 $\beta,~\kappa \setminus \rho_{FM} $
          & $R=2$ & $R=4$   \\ \hline
 -0.5, 0.3150 & 0.122(2)   & 0.057(7)  \\ \hline
 -0.5, 0.3175 & 0.135(4)   & 0.06(1)   \\ \hline
 -0.5, 0.3200 & 0.171(5)   & 0.116(8)  \\ \hline
 -0.5, 0.3225 & 0.235(5)   & 0.28(1)   \\ \hline
 -0.5, 0.3250 & 0.287(3)   & 0.311(7)  \\ \hline
 -0.5, 0.3275 & 0.335(3)   & 0.371(6)  \\ \hline
  0.5, 0.2100 & 0.0574(3)  & 0.0132(6) \\ \hline
  0.5, 0.2125 & 0.0665(8)  & 0.021(1)  \\ \hline
  0.5, 0.2150 & 0.090(2)   & 0.049(2)  \\ \hline
  0.5, 0.2175 & 0.1465(8)  & 0.118(1)  \\ \hline
  0.5, 0.2200 & 0.1913(7)  & 0.171(1)  \\ \hline
  0.5, 0.2225 & 0.2266(5)  & 0.2126(8) \\ \hline
  0.5, 0.2250 & 0.2610(4)  & 0.2504(5) \\ \hline
  2.5, 0.1750 & 0.0463(5)  & 0.0090(9) \\ \hline
  2.5, 0.1775 & 0.058(1)   & 0.020(2)  \\ \hline
  2.5, 0.1800 & 0.103(2)   & 0.071(2)  \\ \hline
  2.5, 0.1825 & 0.1465(7)  & 0.1186(9) \\ \hline
  2.5, 0.1850 & 0.1823(4)  & 0.1574(5) \\ \hline
  2.5, 0.1875 & 0.2149(3)  & 0.1921(3) \\ \hline
\hline
\end{tabular}
\caption{The Fredenhagen--Marcu parameter $\rho_{FM}(R,T)$
         for $\lambda=3.0$ extracted from smeared operators for
         $R=2$ and $R=4$ with $T=R/2$.}
\label{tab:FM_3}
\end{table}

\section{Spectrum Calculations}

To probe the spectrum of the model, we constructed correlation functions
of the following operators:
\begin{eqnarray}
{\cal O}_{S}({\bf x},t)&  = & {\rm Re}
                          \sum_{\mu} \phi_{x}^{*}U_{x,\mu}\phi_{x+\hat\mu} \\
{\cal O}_{V}({\bf x},t)&  = & {\rm Im}
                          \sum_{\mu} \phi_{x}^{*}U_{x,\mu}\phi_{x+\hat\mu} \\
{\cal O}_{P}({\bf x},t)&  = & {\rm Im}
                          \sum_{\Box} U_{\Box}
\end{eqnarray}
with the subscripts $S$, $V$, and $P$ corresponding to the scalar boson,
the vector boson and the photon, and $\mu ,\nu =1,2,3$. These states have
$J^{PC}$ quantum numbers  $0^{++}$, $1^{--}$ and $1^{+-}$ respectively. The
operator ${\cal O}_P$ at  momentum $p = 0 $ has the wrong parity for a
photon state. However, at non--zero momenta, $p > 0 $, parity is no longer a
good quantum number for extended objects like the plaquettes, $U_{\Box}$,
and ${\cal O}_{P}$ will have non--zero overlap with photon states \cite{BP}.

To study the spectrum of the model, we extract the connected  correlations
of a generic operator ${\cal O}$ with quantum numbers  $J^{PC}$ and
momentum $\vec{p}$:

\begin{eqnarray}
C(\tau) &=& \langle {\cal O}(t+\tau ;{\vec p})
{\cal O}^\dagger (t ;{\vec p}) \rangle - \langle {\cal O}(t+\tau ;{\vec p})
          \rangle \langle {\cal O}^\dagger(t ;{\vec p}) \rangle ;
\nonumber \\
\\
{\cal O}(t ;{\vec p}) &=& \sum_{{\vec x}} e^{i{\vec x}{\vec p}} \cdot
 {\cal O}(t ;{\vec x}).
\nonumber
\end{eqnarray}

These correlations are a sum of exponentials exp$(-E_i \tau)$ with $E_i$
the energies of the different states with the given quantum number. We
are interested in the state with the lowest energy, $E_0$, in each channel.
This lowest energy can be approximated by the effective energy
\begin{equation}
E_{eff} = - \ln \frac{C(\tau)}{C(\tau - 1)},
\end{equation}
at large enough $\tau$ with $\tau \ll L_{t}/2$. In practice, the situation
is more complicated as the signals for larger time--slices die away rapidly.
The true energy $E$ could be quite different from the calculated energy
$E_{eff}$ if there is a lot of contamination  from the higher energy
states. In order to maximize the overlap of the  lowest lying state with
our operator and obtain a better signal  for the effective energies, we
used the APE smearing technique and applied  between five and ten smearing
iterations. In our computations, we found  that we were able to get
reliable signals up to $\tau = 2$.  Once the energies $E$ have  been
determined, the masses of the corresponding particles may be  calculated
through the lattice dispersion relation:
\begin{equation}\label{sdisp}
\sinh^{2}\frac{aE}{2} = \sinh^{2}\frac{am}{2} +
                        \sum_{i=1}^{3} \sin^{2} \frac{ap_{i}}{2}.
\end{equation}

The momenta, here, are given by $p_i=2\pi k_i/(aL)$, with  $k_i =
-\frac{1}{2}L +1, \ldots , \frac{1}{2}L$. We shall refer to these momenta
simply as ``$p_i=k_i$'' or as ``$p=k$'', keeping in mind that the momenta
are given in units of $2\pi/(aL)$.

\subsection{Photon}

\begin{figure}
\epsffile{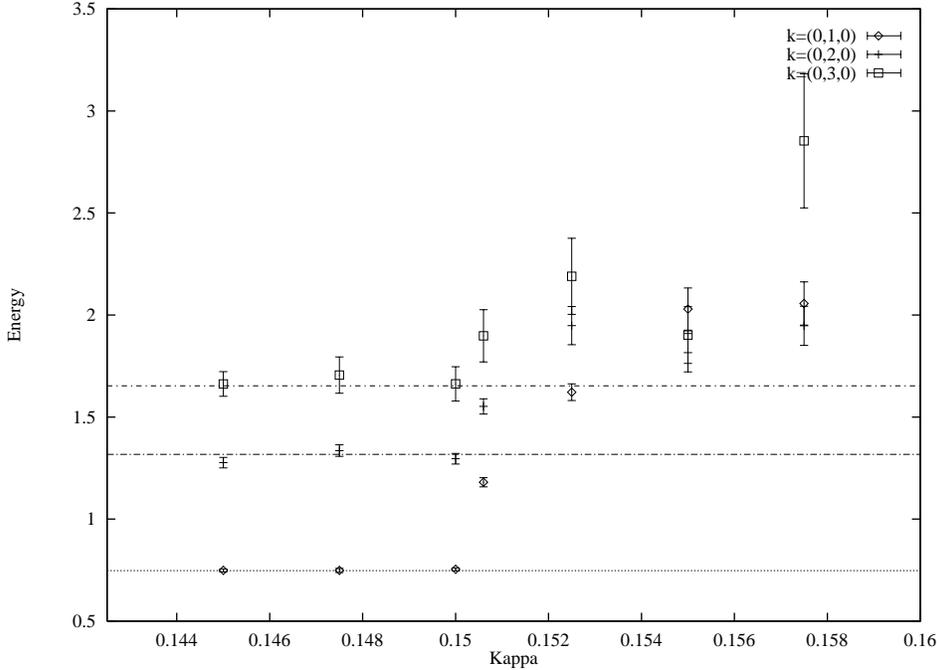}
\caption{Effective Energies of the $1^{+-}$  state at three different
         non--zero momenta as indicated for $\lambda=0.01$ and $\beta=0.5$.
         The lines correspond to the effective energies expected
         from a massless particle.}
\label{phot_1}
\end{figure}

The operator ${\cal O}_{P}$, which at non--zero momentum couples to the
photon, was measured at several non--zero momenta.  The results for the
effective energy at different $\beta$ and $\lambda$ are tabulated in Tables
\ref{tab:m_gam_1} to \ref{tab:m_gam_3}. As an example, we show in Figure
\ref{phot_1} the effective energies for 3 different momenta for
$\lambda=0.01$ and $\beta=0.5$. The lines correspond to the effective
energies that would be expected for a massless particle at these 3 momenta.
As expected, the photon is massless at $\kappa$'s smaller than $\kappa_c$
(in the Coulomb phase) and acquires a mass at $\kappa > \kappa_c$ (in the
Higgs phase). This mass, extracted from the energy with the dispersion
relation (\ref{sdisp}), is given in the last three columns of Tables
\ref{tab:m_gam_1} to \ref{tab:m_gam_3}. We note that the masses extracted
from the dispersion relation are in general agreement with, though somewhat
smaller than, those from the Yukawa fits to the potential (see Tables
\ref{tab:Yuk_fit_2b} and \ref{tab:Yuk_fit_3b}), but they tend to have
larger errors. The dependence of the photon mass on $\kappa$ above the
phase transition (PT) is rather strong for small self--coupling, {\it i.e.},
$\lambda =0.01$, and a discontinuity of the mass can not be excluded. On
the contrary, for large self--coupling ($\lambda =3.0$) the photon mass grows
very slowly above the PT, in agreement with a previous observation in
\cite{EJJLN}.

\begin{figure}
\epsffile{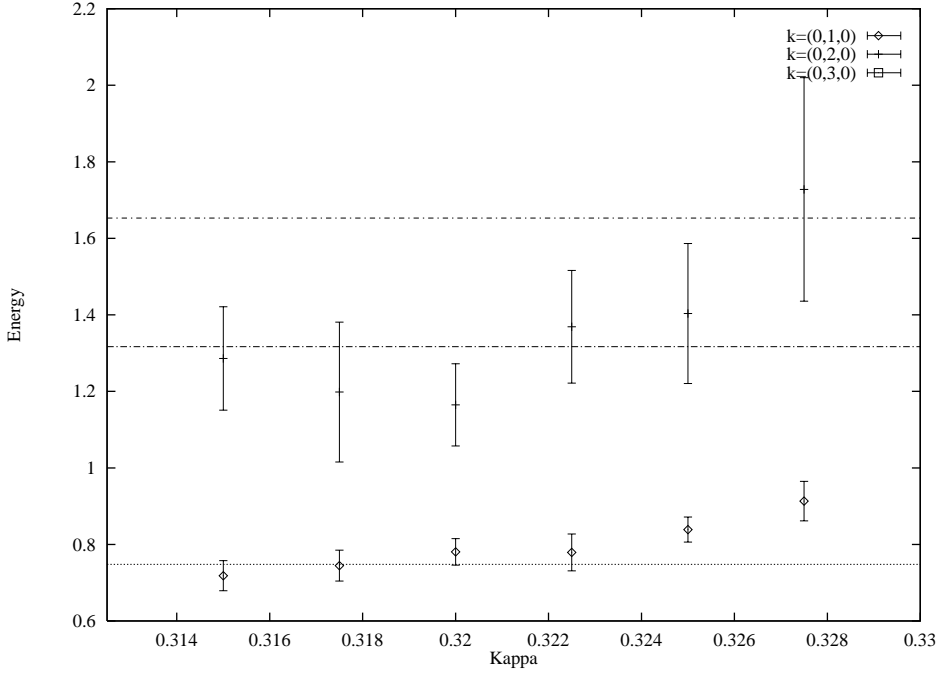}
\caption{Effective Energies of the $1^{+-}$  state at three different
         momenta indicated for $\lambda=3.0$ and $\beta=-0.5$. The
         solid lines correspond to the effective energies expected
         from a massless particle.}
\label{phot_5}
\end{figure}

The study of other order parameters in the previous sections convinced us
that a massless state should exist also at negative $\beta$'s ($\beta >
-0.7$).  This is indeed the case.  As an illustration we show in Figure
\ref{phot_5}  the effective energies for different momenta for
$\lambda=3.0$ and $\beta=-0.5$.

\begin{table}
\centering
\begin{tabular}{||c|c|c|c|c|c|c||}\hline
 $\beta, \kappa$ & $E_{k=1}$ & $E_{k=2}$ & $E_{k=3}$
                 & $m_{k=1}$ & $m_{k=2}$ & $m_{k=3}$ \\ \hline
\hline
 -0.5, 0.1850    & 0.74(2)   & 1.33(5)  & 1.78(8)   & & & \\ \hline
  0.5, 0.1450    & 0.75(1)   & 1.29(4)  & 1.57(8)   & & & \\ \hline
  0.5, 0.1475    & 0.75(1)   & 1.35(4)  & 1.72(7)   & & & \\ \hline
  0.5, 0.1500    & 0.756(6)  & 1.30(5)  & 1.72(11)  & & & \\ \hline
  0.5, 0.1506    & 1.15(3)   & 1.51(5)  & 2.00(13)
                 & 0.91(4)   & 0.84(11) & 1.3(3)    \\ \hline
  0.5, 0.1525    & 1.64(11)  & 2.22(17) & 2.75(49)
                 & 1.51(13)  & 1.97(22) & 2.5(6)    \\ \hline
  0.5, 0.1550    & 2.01(13)  & 1.80(14) & 1.80(20)
                 & 1.93(14)  & 1.38(22) & 0.9(6)    \\ \hline
  0.5, 0.1575    & 2.14(14)  & 1.94(15) & 3.09(97)
                 & 2.07(15)  & 1.59(22) & 2.9(1.2)  \\ \hline \hline
\end{tabular}
\caption{The effective energies and, in the Higgs phase, the
         extracted masses of the photon from unsmeared
         operators at time--slice 2 for $\lambda=0.01$ and for three
         different values of the momentum: $\vec{k}=(0,1,0)$,
         $\vec{k}=(0,2,0)$ and $\vec{k}=(0,3,0)$.
         }
\label{tab:m_gam_1}
\end{table}

\begin{table}
\centering
\begin{tabular}{||c|c|c|c|c|c|c||}\hline
 $\beta, \kappa$ & $E_{k=1}$ & $E_{k=2}$ & $E_{k=3}$
                 & $m_{k=1}$ & $m_{k=2}$ & $m_{k=3}$ \\ \hline
\hline
 -0.5, 0.1850    & 0.72(1)   & 1.32(9)  &           & & & \\ \hline
 -0.5, 0.1875    & 0.68(11)  & 1.30(49) &           & & & \\ \hline
  0.5, 0.1450    & 0.749(8)  & 1.28(3)  & 1.66(6)   & & & \\ \hline
  0.5, 0.1475    & 0.75(1)   & 1.34(3)  & 1.71(9)   & & & \\ \hline
  0.5, 0.1500    & 0.754(7)  & 1.30(3)  & 1.66(8)   & & & \\ \hline
  0.5, 0.1506    & 1.18(2)   & 1.55(4)  & 1.90(13)
                 & 0.95(3)   & 0.93(8)  & 1.1(3)    \\ \hline
  0.5, 0.1525    & 1.62(4)   & 1.95(9)  & 2.19(19)
                 & 1.49(5)   & 1.60(13) & 1.7(3)    \\ \hline
  0.5, 0.1550    & 2.03(10)  & 1.82(9)  & 1.90(14)
                 & 1.95(11)  & 1.41(14) & 1.1(3)    \\ \hline
  0.5, 0.1575    & 2.06(11)  & 1.95(10) & 2.85(33)
                 & 1.98(12)  & 1.60(14) & 2.6(4)    \\ \hline
  2.5, 0.1275    & 0.758(8)  & 1.33(4)  & 1.60(8)   & & & \\ \hline
  2.5, 0.1300    & 0.75(1)   & 1.33(4)  & 1.44(5)   & & & \\ \hline
  2.5, 0.1325    & 0.76(1)   & 1.39(4)  & 1.72(8)
                 & 0.14(6)   & 0.51(14) & 0.6(4)    \\ \hline
  2.5, 0.1350    & 0.84(2)   & 1.30(4)  & 1.63(7)
                 & 0.40(5)   &          &           \\ \hline
  2.5, 0.1375    & 0.90(2)   & 1.40(5)  & 1.78(7)
                 & 0.52(4)   & 0.54(17) & 0.8(2)    \\ \hline \hline
\end{tabular}
\caption{The effective energies and, in the Higgs phase, the
         extracted masses of the photon from smeared
         operators at time--slice 2 for $\lambda=0.01$ and for three
         different values of the momentum: $\vec{k}=(0,1,0)$,
         $\vec{k}=(0,2,0)$ and $\vec{k}=(0,3,0)$.
         }
\label{tab:m_gam_2}
\end{table}

\begin{table}
\centering
\begin{tabular}{||c|c|c|c|c|c|c||}\hline
 $\beta, \kappa$ & $E_{k=1}$ & $E_{k=2}$ & $E_{k=3}$
                 & $m_{k=1}$ & $m_{k=2}$ & $m_{k=3}$ \\ \hline
\hline
 -0.5, 0.3150   & 0.72(4)    & 1.29(14)  & 1.04(1.96) & & & \\ \hline
 -0.5, 0.3175   & 0.74(4)    & 1.20(18)  &            & & & \\ \hline
 -0.5, 0.3200   & 0.78(4)    & 1.16(11)  & 0.7(1.02)
                & 0.23(15)   &           &            \\ \hline
 -0.5, 0.3225   & 0.78(5)    & 1.37(15)  &
                & 0.23(18)   & 0.4(6)    &            \\ \hline
 -0.5, 0.3250   & 0.84(3)    & 1.40(18)  & 1.08(1.43)
                & 0.40(7)    & 0.5(6)    &            \\ \hline
 -0.5, 0.3275   & 0.91(5)    &           & 1.72(29)
                & 0.54(9)    &           &            \\ \hline
  0.5, 0.2100   & 0.740(9)   & 1.30(4)   & 1.51(12)   & & & \\ \hline
  0.5, 0.2125   & 0.75(1)    & 1.25(5)   & 1.49(11)   & & & \\ \hline
  0.5, 0.2150   & 0.76(1)    & 1.28(3)   & 1.71(7)
                & 0.14(6)    &           & 0.5(3)     \\ \hline
  0.5, 0.2175   & 0.79(1)    & 1.32(4)   & 1.74(10)
                & 0.27(3)    &           & 0.7(4)     \\ \hline
  0.5, 0.2200   & 0.82(1)    & 1.36(4)   & 1.79(14)
                & 0.35(3)    & 0.39(18)  & 0.8(4)     \\ \hline
  0.5, 0.2225   & 0.85(1)    & 1.34(5)   & 1.62(12)
                & 0.42(2)    & 0.3(3)    &            \\ \hline
  0.5, 0.2250   & 0.86(1)    & 1.41(5)   & 1.72(10)
                & 0.44(2)    & 0.58(16)  & 0.6(4)     \\ \hline
  2.5, 0.1750   & 0.78(1)    & 1.39(4)   & 1.66(7)    & & & \\ \hline
  2.5, 0.1775   & 0.74(1)    & 1.30(4)   & 1.66(6)    & & & \\ \hline
  2.5, 0.1800   & 0.72(1)    & 1.36(3)   & 1.67(6)    & & & \\ \hline
  2.5, 0.1825   & 0.75(1)    & 1.22(2)   & 1.85(8)
                & 0.06(14)   &           & 1.0(2)     \\ \hline
  2.5, 0.1850   & 0.75(1)    & 1.27(4)   & 1.55(6)
                & 0.06(14)   &           &            \\ \hline
  2.5, 0.1875   & 0.77(1)    & 1.43(5)   & 1.56(7)
                & 0.19(4)    & 0.64(15)  &            \\ \hline \hline
\end{tabular}
\caption{The effective energies and, in the Higgs phase, the
         extracted masses of the photon from smeared
         operators at time--slice 2 for $\lambda=3.0$ and for three
         different values of the momentum: $\vec{k}=(0,1,0)$,
         $\vec{k}=(0,2,0)$ and $\vec{k}=(0,3,0)$.
        }
\label{tab:m_gam_3}
\end{table}

\subsection{Scalar Boson}

It is rather plausible that in the Coulomb phase the scalars $\phi$
interacting through the attractive Coulombic force and the $\phi^4$
coupling can form massive bound--states, {\it bosonium}, analogous to
positronium in normal (fermionic) QED. One of the possible bound--states has
the same quantum numbers, $0^{++}$, as the Higgs boson in the Higgs phase.
The analogy with positronium makes it interesting to subject this state to
a somewhat detailed study. Therefore, we attempted to check its lattice
dispersion relation at different $\beta$ and $\lambda$, including the
negative value $\beta=-0.5$.

We measured at two different momenta, $p=0$ and $p=1$. We were especially
interested to see whether a two free particles dispersion (DR) relation was to
be favored over a single--particle dispersion, indicating a bound--state, in
the Coulomb phase. To investigate this, we calculated the mass of the two
assumed free particles from the $p=0$ energy ($E^{p=0}= 2 m$) and used this
mass to compute the energy $E_{two}^{p=1}$ of the combined two--particle $p=1$
state,
\begin{equation}
E_{two}^{p=1} = m + (2/a) {\rm arsinh} \sqrt{ \sinh^{2}\frac{am}{2} +
 \sin^{2} \frac{\pi}{L} } .
\end{equation}

\begin{figure}
\epsffile{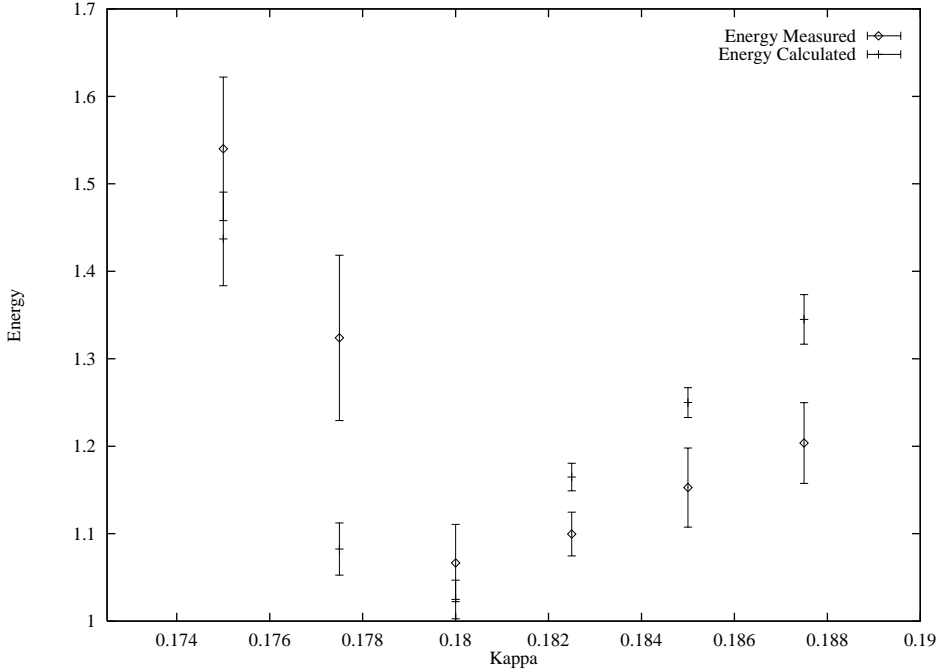}
\caption{Measured and calculated Energies of the $0^{++}$ state at $p=1$
         assuming a two--particle dispersion relation for $\lambda=3.0$
         and $\beta=2.5$.}
\label{sbos_12}
\end{figure}

\begin{figure}
\epsffile{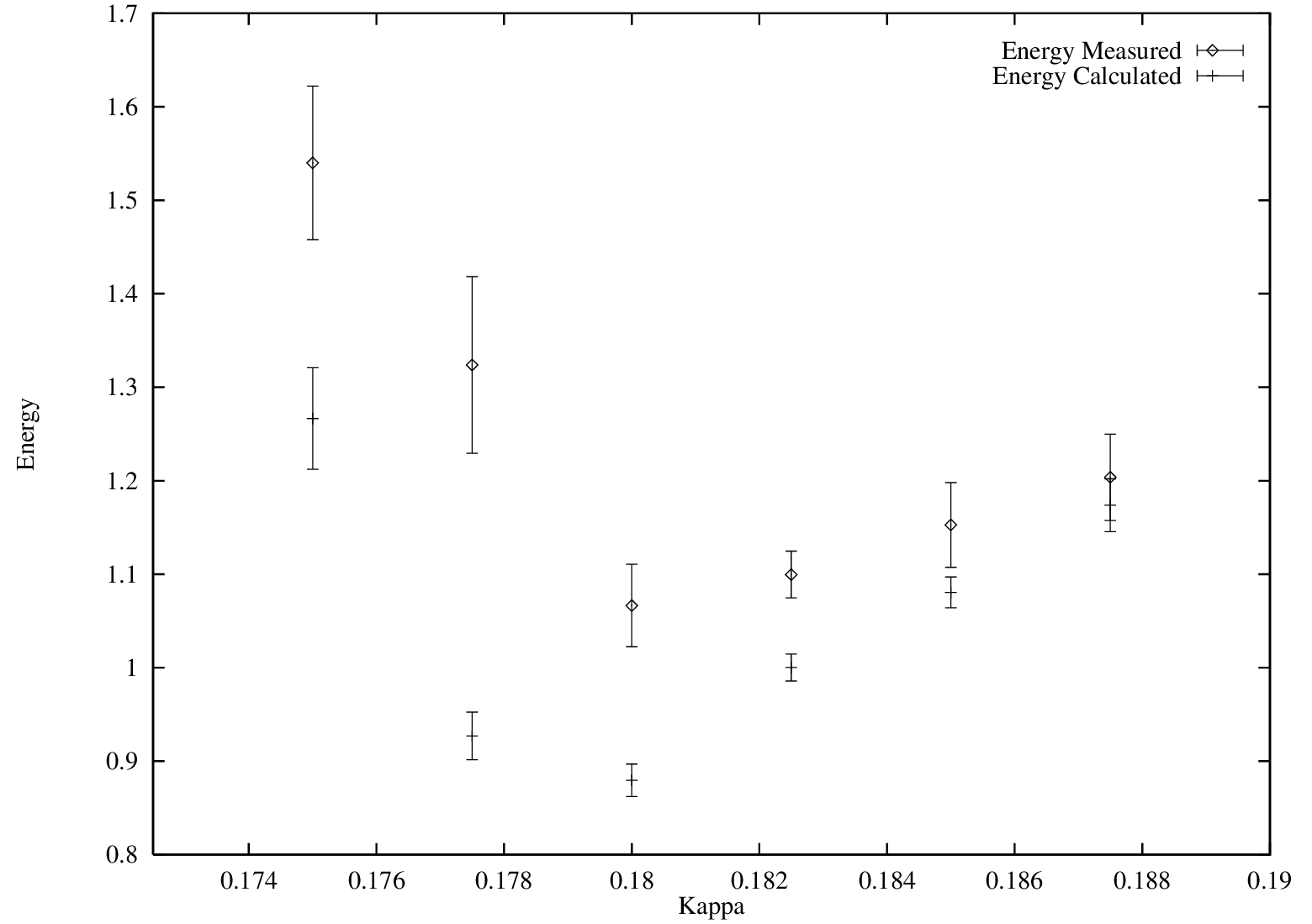}
\caption{Measured and calculated Energies of the $0^{++}$ state at $p=1$
         assuming a single--particle dispersion relation for $\lambda=3.0$
         and $\beta=2.5$.}
\label{sbos_11}
\end{figure}

\begin{figure}
\epsffile{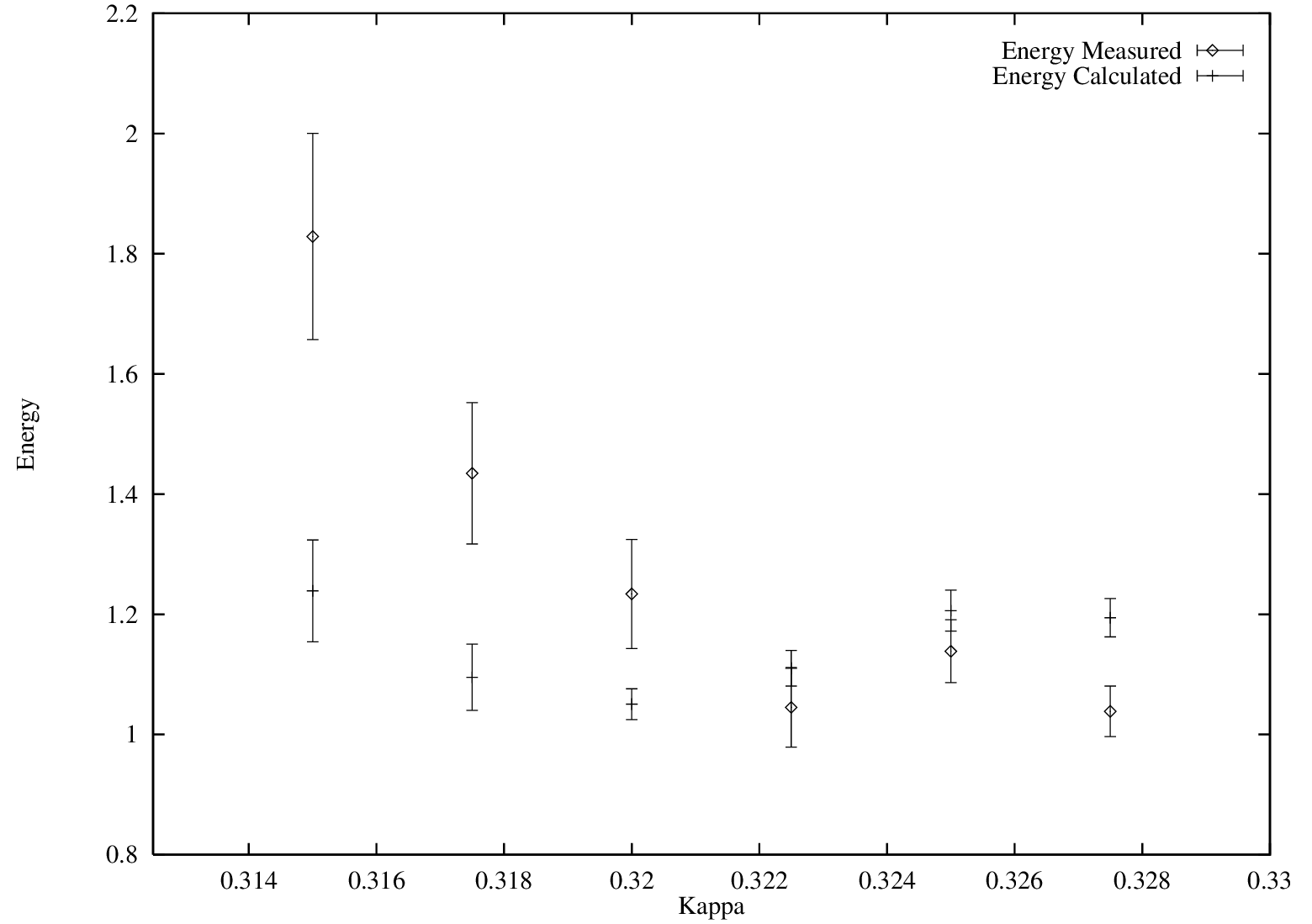}
\caption{Measured and calculated Energies of the $0^{++}$ state at $p=1$
         assuming a two--particle dispersion relation for $\lambda=3.0$
         and $\beta=-0.5$.}
\label{sbos_10}
\end{figure}

\begin{figure}
\epsffile{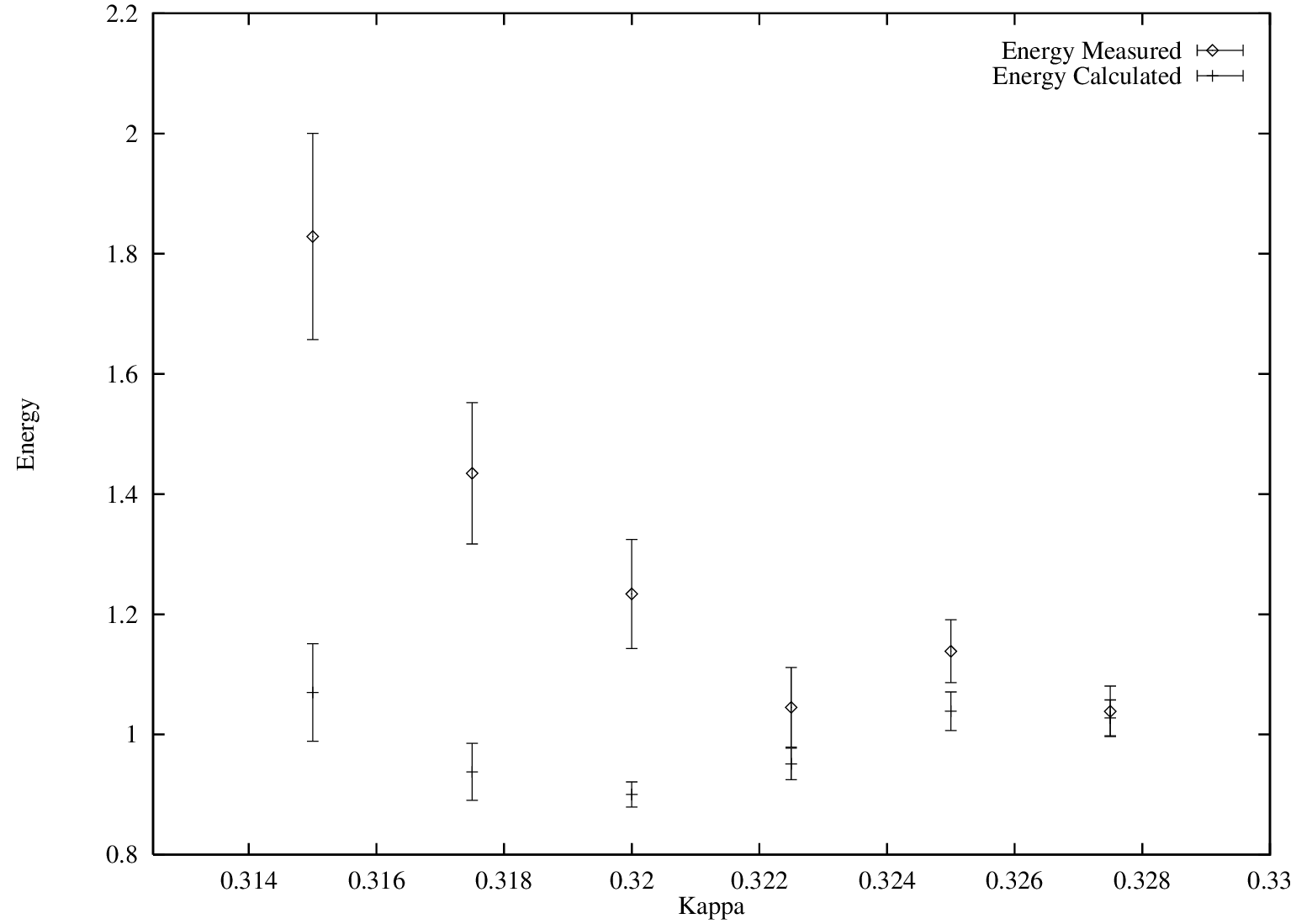}
\caption{Measured and calculated Energies of the $0^{++}$ state at $p=1$
         assuming a single--particle dispersion relation for $\lambda=3.0$
         and $\beta=-0.5$.}
\label{sbos_9}
\end{figure}

\begin{figure}
\epsffile{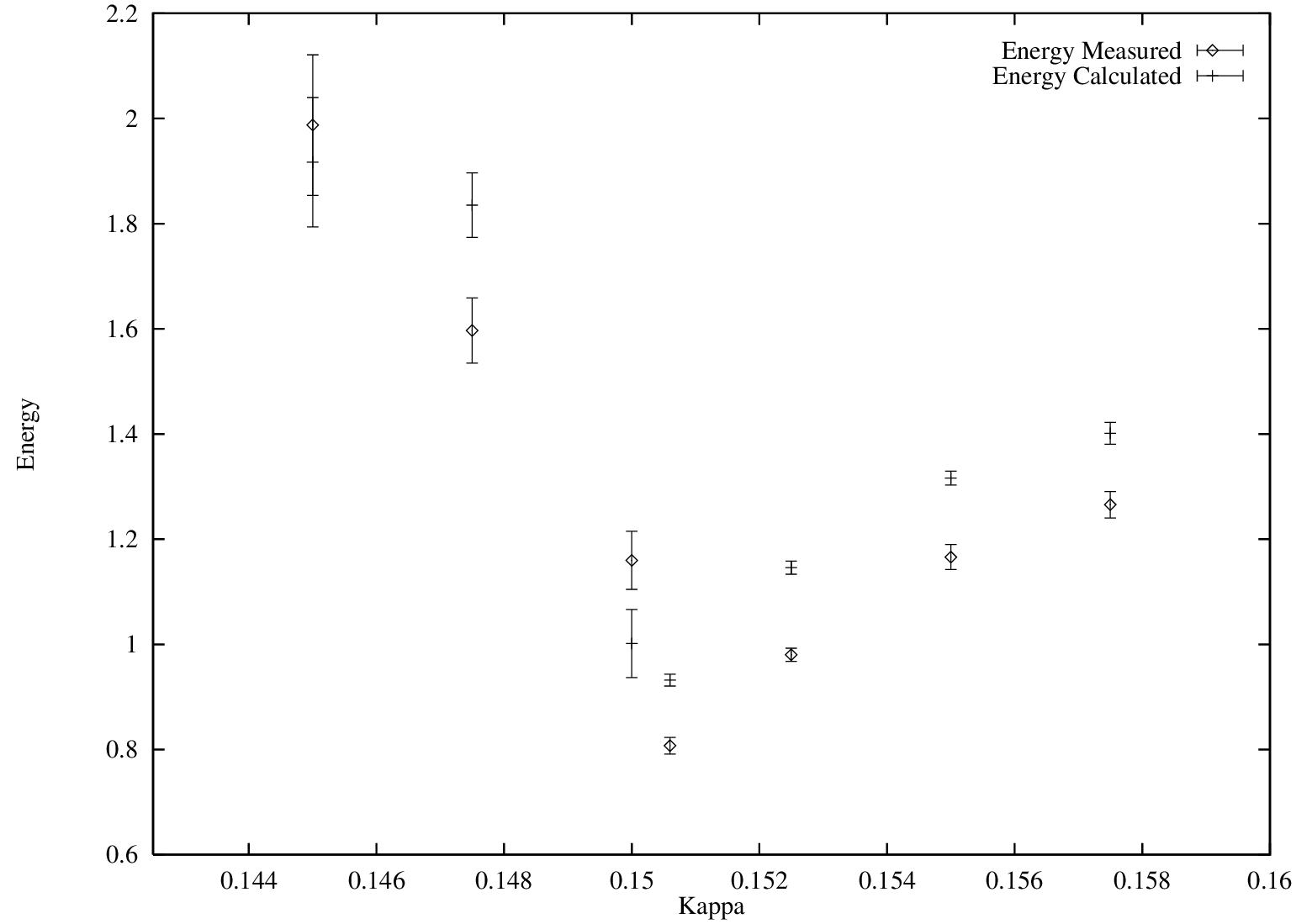}
\caption{Measured and calculated Energies of the $0^{++}$ state at $p=1$
         assuming a two--particle dispersion relation for $\lambda=0.01$
         and $\beta=0.5$.}
\label{sbos_2}
\end{figure}

\begin{figure}
\epsffile{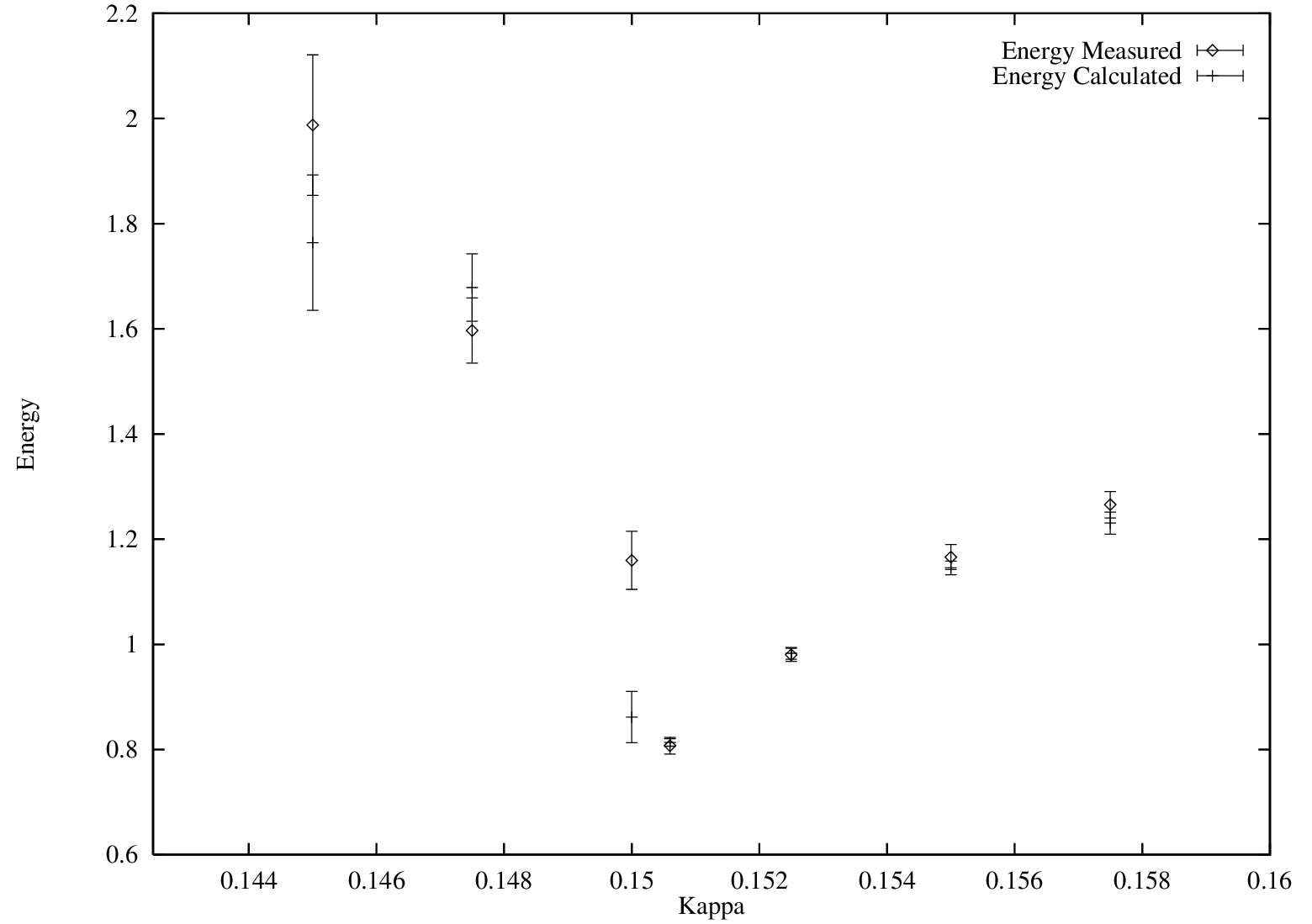}
\caption{Measured and calculated Energies of the $0^{++}$ state at $p=1$
         assuming a single--particle dispersion relation for $\lambda=0.01$
         and $\beta=0.5$.}
\label{sbos_1}
\end{figure}

We then compared the measured energy
$E_{m}^{p=1}$ with the calculated energy $E_{two}^{p=1}$ as well as
with the calculated energy $E_{sing}^{p=1}$ based on the
one--particle dispersion relation.
The results are tabulated in Tables \ref{tab:disp_S_1} to \ref{tab:disp_S_3},
and shown in Figures \ref{sbos_12} to \ref{sbos_1}.

In the Higgs phase the two--particle dispersion relation definitely fails,
while the single--particle dispersion relation is quite consistent for
the two  different momenta $p$ we have chosen. This is, of course,
consistent with the notion that the Higgs boson, in our model, is an
elementary particle.

However, in the Coulomb phase the situation is less clear. Neither the
single nor the two--particle dispersion relations work satisfactorily in the
case of the strong self--interaction ($\lambda =3.0$). As an example, we show
in Figures \ref{sbos_12} and \ref{sbos_11}  the measured and calculated
energies with two--particle DR and one--particle DR, respectively. The
failure of the one--particle DR becomes even more pronounced for negative
$\beta$'s (see Figures \ref{sbos_10} and \ref{sbos_9} ). A possible
explanation could be that, in the Coulomb phase, our energy measurements
are more contaminated by higher excitations.  Unfortunately our signal does
not allow us to go to larger $\tau$ to settle this issue.



In the case of weak self--interaction the one--particle DR looks  somewhat
preferable, as compared to the two--particle DR. In Figures \ref{sbos_2}
and \ref{sbos_1} we show the measured and calculated energies with
two--particle DR and one--particle DR, respectively, for $\beta =0.5$
and $\lambda =0.01$. A qualitatively similar picture was observed for other
$\beta$--values for the same $\lambda$.


\begin{table}
\centering
\begin{tabular}{||c|c|c|c|c||}\hline
 $\beta, \kappa \setminus E_{eff} $ & $E_{meas}^{p=0}$
          & $E_{meas}^{p=1}$ & $E_{two}^{p=1}$ & $E_{sing}^{p=1}$ \\ \hline
\hline
 -0.5, 0.1850   & 1.65(11) & 1.87(14) & 1.91(9)  & 1.76(10) \\ \hline
 -0.5, 0.1850(*)& 0.93(3)  & 1.23(3)  & 1.33(2)  & 1.16(2)  \\ \hline
 -0.5, 0.1875   & 1.06(4)  & 1.35(4)  & 1.43(3)  & 1.26(3)  \\ \hline
 -0.5, 0.1900   & 1.17(12) & 1.48(43) & 1.52(10) & 1.35(10) \\ \hline
  0.5, 0.1450   & 1.84(17) & 2.01(16) & 2.07(15) & 1.93(15) \\ \hline
  0.5, 0.1475   & 1.64(10) & 1.67(7)  & 1.90(9)  & 1.75(9)  \\ \hline
  0.5, 0.1500   & 0.42(9)  & 1.17(5)  & 0.99(6)  & 0.85(4)  \\ \hline
  0.5, 0.1506   & 0.33(2)  & 0.81(2)  & 0.93(1)  & 0.813(7) \\ \hline
  0.5, 0.1525   & 0.66(3)  & 0.97(2)  & 1.14(2)  & 0.98(2)  \\ \hline
  0.5, 0.1550   & 0.91(3)  & 1.15(3)  & 1.32(2)  & 1.15(2)  \\ \hline
  0.5, 0.1575   & 1.07(5)  & 1.25(4)  & 1.44(4)  & 1.27(4)  \\ \hline \hline
\end{tabular}
\caption{The effective energies of the $0^{++}$ scalar boson
         at momentum $p=0$ and $p=1$ from unsmeared operators at time--slice
         2 for $\lambda=0.01$. Tabulated are the effective energies which
         were actually measured and for $p=1$ also those computed from a
         two--particle and from a single--particle dispersion relation for
         the given $E^{p=0}$. The (*) data corresponds to a run from a
         new start, starting from $\kappa=0.1875$.}
\label{tab:disp_S_1}
\end{table}

\begin{table}
\centering
\begin{tabular}{||c|c|c|c|c||}\hline
 $\beta, \kappa \setminus E_{eff} $ & $E_{meas}^{p=0}$
       & $E_{meas}^{p=1}$ & $E_{two}^{p=1}$ & $E_{sing}^{p=1}$ \\ \hline
\hline
 -0.5, 0.1850   & 1.59(9)  & 1.91(15) & 1.86(7)  & 1.71(8)  \\ \hline
 -0.5, 0.1850(*)& 0.93(3)  & 1.24(3)  & 1.33(2)  & 1.16(2)  \\ \hline
 -0.5, 0.1875   & 1.07(4)  & 1.34(4)  & 1.44(3)  & 1.27(3)  \\ \hline
 -0.5, 0.1900   & 1.21(5)  & 1.47(4)  & 1.55(4)  & 1.38(4)  \\ \hline
  0.5, 0.1450   & 1.65(15) & 1.99(13) & 1.92(12) & 1.76(13) \\ \hline
  0.5, 0.1475   & 1.56(7)  & 1.60(6)  & 1.84(6)  & 1.68(6)  \\ \hline
  0.5, 0.1500   & 0.44(11) & 1.16(6)  & 1.00(7)  & 0.86(5)  \\ \hline
  0.5, 0.1506   & 0.33(2)  & 0.81(2)  & 0.93(1)  & 0.813(7) \\ \hline
  0.5, 0.1525   & 0.66(2)  & 0.98(1)  & 1.15(1)  & 0.98(1)  \\ \hline
  0.5, 0.1550   & 0.91(1)  & 1.17(2)  & 1.32(1)  & 1.15(1)  \\ \hline
  0.5, 0.1575   & 1.02(3)  & 1.27(3)  & 1.40(2)  & 1.23(2)  \\ \hline
  2.5, 0.1275   & 1.60(12) & 1.82(17) & 1.87(10) & 1.72(11) \\ \hline
  2.5, 0.1300   & 1.30(10) & 1.56(11) & 1.63(8)  & 1.46(8)  \\ \hline
  2.5, 0.1325   & 0.23(2)  & 0.84(2)  & 0.87(1)  & 0.780(7) \\ \hline
  2.5, 0.1350   & 0.63(2)  & 0.96(2)  & 1.12(1)  & 0.96(1)  \\ \hline
  2.5, 0.1375   & 0.81(1)  & 1.10(2)  & 1.25(1)  & 1.08(1)  \\ \hline \hline
\end{tabular}
\caption{The effective energies of the $0^{++}$ scalar boson
         at momentum $p=0$ and $p=1$ from smeared operators at time--slice
         2 for $\lambda=0.01$. Tabulated are the effective energies which
         were actually measured and for $p=1$ also those computed from a
         two--particle and from a single--particle dispersion relation for
         the given $E^{p=0}$. The (*) data corresponds to a run from a
         new start, starting from $\kappa=0.1875$.}
\label{tab:disp_S_2}
\end{table}

\begin{table}
\centering
\begin{tabular}{||c|c|c|c|c||}\hline
 $\beta, \kappa \setminus E_{eff} $ & $E_{meas}^{p=0}$
          & $E_{meas}^{p=1}$ & $E_{two}^{p=1}$ & $E_{sing}^{p=1}$ \\ \hline
\hline
 -0.5, 0.3150   & 0.80(12) & 1.83(17) & 1.24(9)  & 1.07(8)  \\ \hline
 -0.5, 0.3175   & 0.60(8)  & 1.43(12) & 1.10(5)  & 0.94(5)  \\ \hline
 -0.5, 0.3200   & 0.52(4)  & 1.23(9)  & 1.05(3)  & 0.90(2)  \\ \hline
 -0.5, 0.3225   & 0.61(5)  & 1.05(7)  & 1.11(3)  & 0.95(3)  \\ \hline
 -0.5, 0.3250   & 0.76(5)  & 1.14(5)  & 1.21(3)  & 1.04(3)  \\ \hline
 -0.5, 0.3275   & 0.74(5)  & 1.04(4)  & 1.19(3)  & 1.03(3)  \\ \hline
  0.5, 0.2100   & 1.24(11) & 1.43(9)  & 1.58(9)  & 1.41(9)  \\ \hline
  0.5, 0.2125   & 0.71(4)  & 1.30(8)  & 1.18(3)  & 1.01(2)  \\ \hline
  0.5, 0.2150   & 0.47(2)  & 1.16(4)  & 1.02(1)  & 0.87(1)  \\ \hline
  0.5, 0.2175   & 0.58(2)  & 1.15(5)  & 1.08(1)  & 0.93(1)  \\ \hline
  0.5, 0.2200   & 0.74(3)  & 1.14(3)  & 1.20(2)  & 1.03(2)  \\ \hline
  0.5, 0.2225   & 0.81(3)  & 1.17(4)  & 1.25(2)  & 1.08(2)  \\ \hline
  0.5, 0.2250   & 0.94(3)  & 1.20(4)  & 1.34(2)  & 1.17(2)  \\ \hline
  2.5, 0.1750   & 1.07(6)  & 1.54(8)  & 1.44(5)  & 1.27(5)  \\ \hline
  2.5, 0.1775   & 0.58(5)  & 1.32(10) & 1.08(3)  & 0.93(3)  \\ \hline
  2.5, 0.1800   & 0.49(4)  & 1.07(4)  & 1.03(2)  & 0.88(2)  \\ \hline
  2.5, 0.1825   & 0.69(3)  & 1.10(3)  & 1.17(2)  & 1.00(2)  \\ \hline
  2.5, 0.1850   & 0.81(3)  & 1.15(5)  & 1.25(2)  & 1.08(2)  \\ \hline
  2.5, 0.1875   & 0.94(4)  & 1.20(5)  & 1.35(3)  & 1.17(3)  \\ \hline \hline
\end{tabular}
\caption{The effective energies of the $0^{++}$ scalar boson
         at momentum $p=0$ and $p=1$ from smeared operators at time--slice
         2 for $\lambda=3.0$. Tabulated are the effective energies which
         were actually measured and for $p=1$ also those computed from a
         two--particle and from a single--particle dispersion relation for
         the given $E^{p=0}$.}
\label{tab:disp_S_3}
\end{table}

\pagebreak[4]
\subsection{Vector Boson}

\begin{figure}
\vspace{9.0cm}
\includegraphics{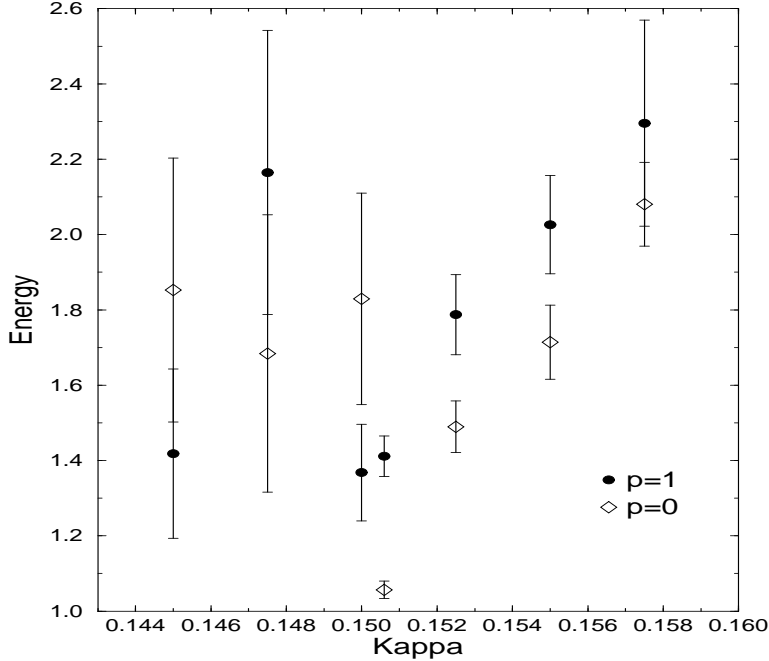}
\vspace{-0.5cm}
\caption{Effective Energies of the $1^{--}$ vector boson at $p=1$
         and $p=0$ for $\lambda=0.01$ and $\beta=0.5$.}
\label{vbos_1}
\end{figure}

Correlations of the operator ${\cal O}_{V}$ allow, in principle, the
measurement of the $1^{--}$ vector boson mass.  This works reasonably well
in the Higgs phase. As an example, we show in Figure \ref{vbos_1} the
effective energies at $p=1$ and $p=0$  for  $\beta =0.5$ and $\lambda =0.01$.
All results are tabulated in Tables \ref{tab:m_V_1} and \ref{tab:m_V_2}.
There is a general tendency for the vector boson mass to decrease, as the
Higgs transition is approached, just as is observed for the scalar boson.
The signal--to--noise ratio is too big to disentangle the dispersion
relations in a way we did for the scalar $~0^{++}~$ boson. Both
one--particle dispersion relation and two--particle dispersion relation
give consistent results within their error bars. Comparing with the masses
extracted from the `photon' operator ${\cal O}_{P}$ at non--zero momenta
(Tables \ref{tab:m_gam_1} to \ref{tab:m_gam_3}) we find rough agreement.
Therefore, the two operators  -- ${\cal O}_V$ and ${\cal O}_P$ -- seem to
couple to the same state in the  Higgs phase, the massive vector boson.
This observation is in accordance with the perturbative analysis of the
spectrum  in the Higgs region, and confirms the early nonperturbative
result \cite{EJJLN} obtained in the case of strong self--coupling ($\lambda
=3.0$).

In the Coulomb phase the situation is quite different from that in the
Higgs phase.  The correlators are very noisy and we do not have sufficient
control of our signals to make, for example, an investigation of the
existence of a bound--state like we attempted for the scalar boson. At
non--zero momentum (there is no massless photon at zero momentum)  the
operator ${\cal O}_{V}$ should couple also to the photon.  However, with
rather large errors the effective energies for both $p=0$ and $p=1$ are
both quite large, $\sim 2/a$, and comparable. The effective energy for
$p=1$, in particular, is much larger than that of the photon, obtained from
the correlations of ${\cal O}_P$. Evidently, the overlap with a photon
state is very small. But due to the large errors on the effective energy,
we can not make convincing arguments in favor of the existence of a vector
bosonium in the Coulomb phase either. To settle the question of the
excitation spectrum in the vector channel in the Coulomb phase would require
much better statistics, or much better operators.

\begin{table}
\centering
\begin{tabular}{||c|c|c|c|c||}\hline
 $\beta, \kappa$ & $E^{norm}_{eff}(p=0)$ & $E^{norm}_{eff}(p=1)$
                 & $E^{smear}_{eff}(p=0)$ & $E^{smear}_{eff}(p=1)$ \\ \hline
\hline
 0.5, 0.1450    &           &           & 1.85(35)  & 1.42(22)  \\ \hline
 0.5, 0.1475    & 1.48(36)  &           & 1.68(37)  & 2.16(38)  \\ \hline
 0.5, 0.1500    & 1.79(31)  & 1.34(15)  & 1.83(28)  & 1.37(13)  \\ \hline
 0.5, 0.1506    & 1.13(3)   & 1.28(36)  & 1.06(2)   & 1.41(6)   \\ \hline
 0.5, 0.1525    & 1.44(11)  & 1.75(11)  & 1.49(7)   & 1.79(11)  \\ \hline
 0.5, 0.1550    & 1.77(10)  & 1.90(12)  & 1.71(10)  & 2.03(13)  \\ \hline
 0.5, 0.1575    & 2.10(11)  & 2.25(31)  & 2.08(11)  & 2.30(27)  \\ \hline
 2.5, 0.1275    &           &           & 2.22(87)  & 2.6(1.6)  \\ \hline
 2.5, 0.1300    &           &           & 1.45(21)  & 2.6(1.0)  \\ \hline
 2.5, 0.1325    &           &           & 0.47(4)   &           \\ \hline
 2.5, 0.1350    &           &           & 0.42(1)   & 1.13(12)  \\ \hline
 2.5, 0.1375    &           &           & 0.56(1)   & 0.90(6)   \\ \hline \hline
\end{tabular}
\caption{The effective energies for the $1^{--}$ vector boson
         from unsmeared (normal) and smeared operators at time--slice 2 for
         $\lambda=0.01$. There was no signal for entries left blank,
         and none at all for $\beta=-0.5$.}
\label{tab:m_V_1}
\end{table}

\begin{table}
\centering
\begin{tabular}{||c|c|c||}\hline
 $\beta, \kappa$ & $E_{eff}(p=0)$ & $E_{eff}(p=1)$ \\ \hline
\hline
 -0.5, 0.3150   & 2.29(90)  & 1.18(23)  \\ \hline
 -0.5, 0.3175   & 1.69(55)  & 1.46(45)  \\ \hline
 -0.5, 0.3200   & 1.09(26)  & 1.31(41)  \\ \hline
 -0.5, 0.3225   & 1.11(26)  & 1.36(37)  \\ \hline
 -0.5, 0.3250   & 1.14(16)  &           \\ \hline
 -0.5, 0.3275   & 0.83(17)  & 2.0(1.5)  \\ \hline
  0.5, 0.2100   &           &           \\ \hline
  0.5, 0.2125   & 1.48(21)  & 1.49(23)  \\ \hline
  0.5, 0.2150   & 0.91(18)  & 1.65(39)  \\ \hline
  0.5, 0.2175   & 0.86(8)   &           \\ \hline
  0.5, 0.2200   & 0.68(5)   & 1.52(43)  \\ \hline
  0.5, 0.2225   & 0.68(5)   & 1.77(52)  \\ \hline
  0.5, 0.2250   & 0.66(3)   & 1.84(37)  \\ \hline
  2.5, 0.1750   & 1.89(44)  & 2.8(1.6)  \\ \hline
  2.5, 0.1775   & 1.49(26)  & 2.25(62)  \\ \hline
  2.5, 0.1800   & 0.90(19)  &           \\ \hline
  2.5, 0.1825   & 0.51(5)   & 2.7(1.2)  \\ \hline
  2.5, 0.1850   & 0.52(7)   & 2.3(1.3)  \\ \hline
  2.5, 0.1875   & 0.36(4)   & 2.4(1.2)  \\ \hline \hline
\end{tabular}
\caption{The effective energies for the $1^{--}$ vector boson
         from smeared operators at time--slice 2 for $\lambda=3.0$.
         There was no signal for entries left blank.}
\label{tab:m_V_2}
\end{table}

\section{Conclusions}

We have made a thorough study of the compact $U(1)$ lattice gauge--Higgs
model with completely suppressed monopoles. We studied it both for
comparatively small self--coupling ($\lambda =0.01$), as well as in the
strong self--coupling limit ($\lambda =\infty$ and $\lambda =3.0$). For
strong self--coupling most computations were performed with $\lambda =3.0$,
permitting us to compare our results with those of a previous study
\cite{EGJJKN,EJJLN} for the standard action. The computations were
performed for three different values of the gauge coupling $\beta$: one at
relatively weak gauge coupling, $\beta=2.5$, one at stronger coupling,
$\beta=0.5$, and one at negative $\beta$, corresponding to imaginary bare
gauge coupling.

We first mapped out the phase diagram of our model with monopoles
suppressed. Contrary to the model with monopoles, it does not exhibit a
confinement phase for strong gauge coupling. Nothing special seems to
happen at $\beta =0$, and both Higgs and Coulomb phases continue to
negative $\beta$ until about $\beta=-0.7$, where a first order transition
into a frustrated phase seems to occur. The Higgs transition separating
Coulomb and Higgs phase at positive $\beta$ is clearly of first order for
$\lambda =0.01$, similar to what happens in the model with the standard
action \cite{mitrju}. Following Coleman and Weinberg \cite{cw}, it could be
interpreted as a radiatively induced spontaneous symmetry breaking. At
strong self--coupling the Higgs transition becomes weaker and its order
harder to determine. From our data we can not exclude that the transition
becomes second order for sufficiently large $\lambda$'s, similar to the
non--compact case \cite{bfkk}.

We measured $T$--dependent potentials from Wilson loops. In the Coulomb
phase they could be well fit with a $T$--dependent Coulomb potential
(extracted from perturbative Wilson loops), even at the negative $\beta$
probed, at least when the potential was extracted from smeared Wilson
loops. The renormalized gauge coupling $g_R^2$ grows with decreasing
$\beta$, and continues to rise rapidly at negative $\beta$.  The
suppression of monopoles therefore does not seem to exclude a
strong--coupling QED, but shifts it to the (unphysical) region $g^2_{bare}
<0$. Since there the bare coupling is imaginary the transfer matrix is no
longer positive and it is thus not clear whether the model makes sense
physically. However, we would like to stress again that we detected no
indication of a phase transition between the positive and negative $\beta$
regions.

In the Higgs phase we fitted the $T$--dependent potentials to
$T$--dependent Yukawa potentials. This worked well for positive $\beta$,
and the extracted vector boson masses agreed reasonably with those obtained
from conventional correlation functions. At negative $\beta$ the  potential
looked rather strange, and no good fits could be obtained.

To study the spectrum of the model we calculated the correlation functions
corresponding to the photon (at non--zero momentum), the scalar boson with
quantum numbers $0^{++}$ and the vector boson with $1^{--}$.

Our results conformed to our expectations, in particular for the photon. In
the Coulomb phase it was found to be massless (even at negative $\beta$),
and its energy fitted well the lattice dispersion relation. At $\kappa
>\kappa_c$ (in the Higgs phase) the photon acquired a mass. The estimation
of $\kappa_c$ from the behavior of the photon is in a good agreement with
that obtained from the Fredenhagen--Marcu order parameter.

The measurements were more difficult for the $0^{++}$ and $1^{--}$ bosons,
especially in the Coulomb phase. Both are reasonably well defined in the
Higgs phase, as one would expect. Determination of the dispersion relation
gave reasonably good evidence that the $0^{++}$ boson, the Higgs boson, is
an elementary particle in the Higgs phase. In the Coulomb phase, however,
the one--particle dispersion relation does not look very convincing, just
like in the model with the standard action \cite{EGJJKN,EJJLN}. Within our
large errors a two--particle dispersion relation, appropriate for a weakly
bound state, did not work much better. Therefore, we do not think that the
problem of the existence of {\it bosonium}, a bound--state between two scalar
particles, in the Coulomb phase is settled. Much better statistics and/or
better operators are needed to settle this question.

\section*{Acknowledgements}

The work of UMH was partly supported by the DOE under grants
\#~DE-FG05-85ER250000 and \#~DE-FG05-92ER40742. The computations were
performed on the cluster of VMS alpha workstations at SCRI and on
CONVEX mainframes at Humboldt-Universit\"at and DESY-Zeuthen.

\end{document}